\documentclass[preprint,12pt]{elsarticle}

\usepackage{amssymb}
\usepackage[cmex10]{amsmath}

\biboptions{numbers,sort&compress}
\usepackage{lscape}
\usepackage[utf8]{inputenc}

\usepackage{graphicx}
\usepackage{booktabs}
\usepackage{lipsum}% just to generate filler text
\usepackage{float}
\usepackage[hidelinks]{hyperref}
\usepackage{xcolor}
\definecolor{Darkgreen}{rgb}{0,0.4,0}
\definecolor{Darkred}{rgb}{0.6,0,0}
\definecolor{Darkblue}{rgb}{0,0,0.4}
\hypersetup{colorlinks = true,
   linkcolor = Darkred,
   citecolor = Darkblue,
   urlcolor = Darkblue,
}

\journal{Renewable \& Sustainable Energy Reviews}

\begin{document}

\begin{frontmatter}



\title{Peer-to-peer and community-based markets: A comprehensive review}

\author[elma]{Tiago Sousa\corref{co}}
\ead{tsousa@elektro.dtu.dk}
\cortext[co]{Corresponding author}

\author[inesc]{Tiago Soares}
\ead{tasoares@inesctec.pt}

\author[elma]{Pierre Pinson}
\ead{ppin@elektro.dtu.dk}

\author[elma]{Fabio Moret}
\ead{fmoret@elektro.dtu.dk}

\author[satie]{Thomas Baroche}
\ead{thomas.baroche@ens-rennes.fr}

\author[elma]{Etienne Sorin}
\ead{egsorin@elektro.dtu.dk}

\address[elma]{Department of Electrical Engineering, Technical University of Denmark, 2800 Kongens Lyngby, Denmark}
\address[inesc]{INESC Technology and Science (INESC TEC), 4200-465 Porto, Portugal}
\address[satie]{SATIE Laboratory located at Ecole Normale Superieure de Rennes, France}

\begin{abstract}
The advent of more proactive consumers, the so-called "prosumers", with production and storage capabilities, is empowering the consumers and bringing new opportunities and challenges to the operation of power systems in a market environment. Recently, a novel proposal for the design and operation of electricity markets has emerged: these so-called peer-to-peer (P2P) electricity markets conceptually allow the prosumers to directly share their electrical energy and investment. Such P2P markets rely on a consumer-centric and bottom-up perspective by giving the opportunity to consumers to freely choose the way they are to source their electric energy. A community can also be formed by prosumers who want to collaborate, or in terms of operational energy management. This paper contributes with an overview of these new P2P markets that starts with the motivation, challenges, market designs moving to the potential future developments in this field, providing recommendations while considering a test-case.
\end{abstract}


\end{frontmatter}


% Introduction
\section{Introduction}\label{Intro}
% DERs and ICT
The continuous integration of Distributed Energy Resources (DERs) \cite{Bussar2016}, e.g., from rooftop solar panels, storage and control devices, along with the advance in Information and Communication Technology (ICT) devices \cite{Saad2014} are inducing a transformation of a share of electricity consumers into prosumers. Prosumers undertake a proactive behavior by managing their consumption, production and energy storage \cite{Zafar2018, Eurelectric2015}, while traditional consumers assume a passive behavior when it comes to their energy consumption\footnote{Prosumer covers a wider-spectrum of consumers with different assets and behaviors, and therefore this term is adopted in the rest of the paper}. Besides technology-based advances, the collaborative economy principle is influencing how prosumers perceive electric energy \cite{Eurelectric2015}, from their increasingly engagement with energy community initiatives \cite{Schoor2015}, to their desire of more flexibility on choosing who they are going to exchange energy \cite{Bertsch2016}. 

% Collaborative economy principle
Lately, the emergence of this principle is changing the way society trades goods and services \cite{Selloni2017}. One can see it as a variety of players, with equal access to a common resource and goal of sharing it through a wealth of cooperating infrastructures, which is opposite to the traditional economic principle represented by players with individual goals, and with the resulting equilibrium occurring when all individual goals are satisfied\footnote{For a deeper understanding about collaborative economy and its different classes with their own characteristics please refer to \cite{Raworth2017, Bollier2015, Pais2015}}. For instance, there are prominent examples of this paradigm change in our daily lives, with various degrees of collaboration\footnote{An interested reader may consult an online repository of collaborative platforms at \url{http://meshing.it/}}, such as BlaBlaCar\footnote{Carpooling and share the cost of the journey - \url{https://www.blablacar.com/}}, Taskrabbit\footnote{Collaboration with housekeeping task - \url{https://www.taskrabbit.com/}} and Turo\footnote{Collaboration with renting cars - \url{https://turo.com/}}.

% Transition to consumer-centric market
Although power systems are evolving to a more decentralized management, electricity markets still perform resource allocation and pricing based on the conventional hierarchical and top-down approach \cite{Hu2017} of power system management, which makes prosumers behave as passive receivers. Reorganizing electricity markets within decentralized management and collaborative principle will instead allow for a bottom-up approach that would empower prosumers \cite{Peng2017}. This may then dynamically influence the market through implementation of prosumers' preferences \cite{Faber2014}, for instance renewable type, CO$_2$ emissions and localized energy. Consequently, this alternative market organization is generically named as {\it consumer-centric electricity markets}, while nearly 20 years ago this visionary concept was mainly seen as a point of academic discussion to weight advantages and drawbacks of centralized vs. decentralized market structures \cite{Wu1995, Wu1999}.

% Using P2P and community-based structures
This consumer-centric market view relies on Peer-to-Peer (P2P) and community-based structures \cite{Giotitsas2015, Morstyn2018b}. P2P\footnote{First application was in computer science to distribute data, and interested readers may consult \cite{Schollmeier2001, Singh2001, Kant2002, Oram2001, Aberer2002, Benkler2006, Vu2010, Peleg2011}} defines a decentralized structure where all peers cooperate with what they have available for commons-based producing, trading or distributing a good or service. P2P and community-based concepts have strongly been applied under a collaborative economy principle\footnote{Examples as Turo and Taskrabbit use P2P structures} as the structure that facilitates the exchange of commons amongst all agents (or peers) \cite{Kostakis2016, Einav2016}, and they are rather distinct from the centralized structures seen in some traditional economic sectors. In fact, the first reference proposing P2P concept for power systems can be traced back to 2007 \cite{Beitollahi2007}, as well as there are practical examples using P2P to share energy in local areas, like the iconic case of the Brooklyn microgrid project \cite{Mengelkamp2017}. However, most countries still prohibit direct energy exchanges between prosumers, though first attempts can be seen in \cite{EU2016, French2017} to adapt regulation. These initial steps make us believe in this novel vision of electricity markets as a possible future.

% Paper contribution
The goal of this paper is to provide a comprehensive understanding of relevant consumer-centric electricity markets\footnote{For simplicity, P2P markets to generically refer as consumer-centric electricity markets}. A test-case is included with data to encourage reproducibility and benchmarking in future research on P2P markets. Recent studies \cite{Morstyn2018b, Jogunola2017, Tushar2018} focused more on the market prospects and technological aspects related to integrating prosumers. In contrast, our goal is to enable academic and industrial communities to better understand all aspects related to this transition, to explain how and why they are emerging, and to be capable of proposing new market structures and business models. Although the importance of prosumers is not undermined in our study, the authors look at P2P markets from a wider perspective that includes all involved agents in the power system. Indeed, a peer is defined as anyone owning or operating an asset or group of assets (e.g., production, consumption, storage). More generally, all potential active agents in the market can be seen as peers. As is the case today, some of the agents in the wholesale market do not own and operate assets, instead they trade on behalf of others or possibly are involved in arbitrage and virtual bidding.

% Paper structure
The paper is organized as follows, while explaining the review methodology followed by the authors in \ref{research_method}. The premises that support P2P markets, as well as research projects and companies, are discussed in Section \ref{P2P_status}. An analysis of the different P2P market structures, including a description of suitable optimization techniques for negotiation and market clearing, are described in Section \ref{P2P_market}. Section \ref{Opportunities_challenges} identifies the opportunities and challenges that can arise when adopting P2P markets. A benchmark test case illustrating the application of P2P markets is presented in Section \ref{Test_case}, which will be available for others to use and test future work in this field. Finally, Section \ref{Conc_recom} gathers a set of conclusions, as well as recommendations for future work.

% P2P history and current status
\section{Premises leading towards peer-to-peer markets}\label{P2P_status}
This section explains the premises in terms of economic (bilateral contracts) and technology (microgrids) that enable the emergence of P2P markets in the energy sector. Then, we also address the current Research \& Development (R{\&}D) projects and start-up companies in this field.

\subsection{Bilateral contracts and microgrids}\label{Bila_cont_and_microgrids}
% Bilateral contracts
Bilateral contracts were introduced with the aim of increasing competition in electricity markets \cite{Bower2000, Hausman2008}. A bilateral contract is an agreement between two parties (buyer and seller) to exchange electric energy, generation capacity rights or related products for a specified period of time, as well as at an agreed price. Under the concept of decentralized systems, Gui \textit{et al.} \cite{Gui2017} analysed the impact of bilateral agreements on community microgrids. This study states that in a microgrid context, the service provider and the consumers will have a strong relationship that can affect incentives and governance models. Wu and Varaiya \cite{Wu1995, Wu1999} proposed in the nineties a coordinated multi-bilateral trading model as a credible alternative to the pool structure used in the wholesale electricity markets. This model was originally proposed for large players that operate in the market and not for small-scale DERs, but it establishes the premises for a P2P market. Indeed, in its simplest form, a P2P market implies multi-bilateral agreements between agents.

% Microgrids
Microgrids are generally accepted as a low voltage distribution grid comprising DERs that can be operated in islanding and grid connected mode \cite{Hirsch2018}. Thus, Distributed System Operators (DSOs) ought to rethink their grid management practice to address this technology transition by adopting novel concepts as addressed in\footnote{e.g. engaging in active distribution network management} \cite{Eurelectric2013, Zhao2014}. In this context, microgrids are relevant in assisting DSOs \cite{Palizban2014}. The deployment of microgrids brings infrastructures and technologies in the domains of monitoring, communication and control that are important enablers for P2P markets. The works in \cite{Liu2017, Ilic2012, Mengelkamp2017} have substantially explored the technological promises brought by microgrids to propose P2P market solutions.

\subsection{Research projects and companies with a relation to P2P markets}\label{P2P_projects_and_companies}
In recent years, a number of R{\&}D projects have been carried out with twofold purposes, as discussed in \cite{Zhang2017}: (i) working on the market design and business models for P2P markets; (ii) implementing local control and ICT platforms for prosumers and microgrids. Table \ref{P2P_proj_overview} summarizes the the R{\&}D projects involved in P2P markets.

\begin{table}[htb]
\centering
\caption{Overview of the R{\&}D projects.}
\label{P2P_proj_overview}
\resizebox{\textwidth}{!}{%
\begin{tabular}{llllll}
\hline
Project name & Country/Region & Starting year & Focus level & Outcomes & Classification \\ \hline
P2P-SmartTest & \begin{tabular}[c]{@{}l@{}}Europe (Finland,\\ United Kingdom,\\ Spain, Belgium)\end{tabular} & \begin{tabular}[c]{@{}l@{}}2015 \\ (ongoing)\end{tabular} & \begin{tabular}[c]{@{}l@{}}Distribution \\ grid level\end{tabular} & \begin{tabular}[c]{@{}l@{}}Advanced control \\ and ICT for P2P \\ energy market\end{tabular} & \begin{tabular}[c]{@{}l@{}}Local control \\ and ICT; \\ Market design\end{tabular} \\ \hline
EMPOWER & \begin{tabular}[c]{@{}l@{}}Europe (Norway;\\ Switzerland, Spain,\\ Malta, Germany)\end{tabular} & \begin{tabular}[c]{@{}l@{}}2015\\ (ongoing)\end{tabular} & \begin{tabular}[c]{@{}l@{}}Distribution\\ grid level\end{tabular} & \begin{tabular}[c]{@{}l@{}}Architecture and ICT\\ solutions for provider\\ in local market\end{tabular} & \begin{tabular}[c]{@{}l@{}}Local control\\ and ICT\end{tabular} \\ \hline
NRGcoin & \begin{tabular}[c]{@{}l@{}}Europe (Belgium,\\ Spain)\end{tabular} & 2013 (finish) & \begin{tabular}[c]{@{}l@{}}Consumer/\\ prosumer\end{tabular} & \begin{tabular}[c]{@{}l@{}}P2P wholesale\\ trading platform\end{tabular} & Market design \\ \hline
Enerchain & Europe & \begin{tabular}[c]{@{}l@{}}2017 \\ (ongoing)\end{tabular} & \begin{tabular}[c]{@{}l@{}}Wholesale\\ market\end{tabular} & \begin{tabular}[c]{@{}l@{}}P2P wholesale\\ trading platform\end{tabular} & Market design \\ \hline
\begin{tabular}[c]{@{}l@{}}Community\\ First! Village\end{tabular} & USA & \begin{tabular}[c]{@{}l@{}}2015\\ (ongoing)\end{tabular} & \begin{tabular}[c]{@{}l@{}}Consumer/\\ prosumer\end{tabular} & \begin{tabular}[c]{@{}l@{}}Build self-sustained\\ community for\\ homeless\end{tabular} &  \begin{tabular}[c]{@{}l@{}}Local control\\ and ICT\end{tabular} \\ \hline
\begin{tabular}[c]{@{}l@{}}PeerEnergy\\ Cloud\end{tabular} & Germany & 2012 (finish) & \begin{tabular}[c]{@{}l@{}}Consumer/\\ prosumer\end{tabular} & \begin{tabular}[c]{@{}l@{}}Cloud-based energy\\ trading for excessive\\ production\end{tabular} & \begin{tabular}[c]{@{}l@{}}Local control\\ and ICT\end{tabular} \\ \hline
Smart Watts & Germany & 2011 (finish) & \begin{tabular}[c]{@{}l@{}}Consumer/\\ prosumer\end{tabular} & \begin{tabular}[c]{@{}l@{}}ICT to control \\ consumption in a\\ secure manner\end{tabular} & \begin{tabular}[c]{@{}l@{}}Local control\\ and ICT\end{tabular} \\ \hline
NOBEL & \begin{tabular}[c]{@{}l@{}}Europe (Germany,\\ Spain, Greece,\\ Sweden, Spain)\end{tabular} & 2012 (finish) & \begin{tabular}[c]{@{}l@{}}Consumer/\\ prosumer\end{tabular} & \begin{tabular}[c]{@{}l@{}}ICT for energy\\ brokerage system\\ with consumers\end{tabular} & \begin{tabular}[c]{@{}l@{}}Local control\\ and ICT\end{tabular} \\ \hline
Energy Collective & Denmark & \begin{tabular}[c]{@{}l@{}}2016\\ (ongoing)\end{tabular} & \begin{tabular}[c]{@{}l@{}}Consumer/\\ prosumer\end{tabular} & \begin{tabular}[c]{@{}l@{}}Deployment of local \\ P2P markets in \\ Denmark\end{tabular} & Market design \\ \hline
P2P3M & \begin{tabular}[c]{@{}l@{}}Europe (United \\ kingdom), \\ Asia (South Korea)\end{tabular} & \begin{tabular}[c]{@{}l@{}}2016\\ (ongoing)\end{tabular} & \begin{tabular}[c]{@{}l@{}}Consumer/\\ prosumer\end{tabular} & \begin{tabular}[c]{@{}l@{}}Prototype P2P energy\\ trading/sharing \\ platform\end{tabular} & Market design \\ \hline
\end{tabular}%
}
\end{table}

% Briefly describe ICT control
In the load control and ICT level, EMPOWER\footnote{\url{http://empowerh2020.eu/}} developed a real-time platform \cite{Massague2017} based on cloud technology to execute the metering and trading within a local community. The P2P-SmartTest project \footnote{\url{http://www.p2psmartest-h2020.eu/}} is exploring distributed control with advanced ICT to enable local markets on a distribution grid. The project points out that, in terms of control, the main challenges in distribution grids are the low inertia, uncertainty and stability issues.

% Briefly describe market design
In terms of market design proposals, the Enerchain project \footnote{\url{https://enerchain.ponton.de}} intends to develop a P2P trading platform to complement, or replace, the wholesale electricity market\footnote{\url{https://enerchain.ponton.de/index.php/21-enerchain-p2p-trading-project}}. On the other hand, NRGcoin\footnote{\url{http://nrgcoin.org/}} aims to develop a virtual currency \cite{Mihaylov2014} based on blockchain and smart contracts for small prosumers trading in P2P markets. The Energy Collective project\footnote{\url{http://the-energy-collective-project.com/}} investigates P2P market designs for local energy communities. An educational APP\footnote{\url{https://p2psystems.shinyapps.io/ShinyApp_Project/}} has been developed to educate a broader audience about P2P markets and their promises.

% Describe companies
At the same time, start-ups have emerged from R{\&}D projects \cite{Hasse2017, Johnston2017} to address P2P energy trading by focusing on the following business areas: (i) P2P exchange of energy surplus, where prosumers can exchange the energy surplus with their neighbours, for example through the companies - LO3 Energy\footnote{\url{http://lo3energy.com/transactive-grid/}}, SonnenCommunity\footnote{\url{https://microsite.sonnenbatterie.de/en/sonnenCommunity}}, Hive Power\footnote{\url{https://www.hivepower.tech/}}, OneUp\footnote{\url{https://www.oneup.company/}}, Power Ledger\footnote{\url{https://tge.powerledger.io/media/Power-Ledger-Whitepaper-v3.pdf}}; (ii) Energy provision/matching, where prosumers can directly choose local renewable generation, for example  through the companies - Vandebron\footnote{\url{https://vandebron.nl}}, Electron\footnote{\url{http://www.electron.org.uk/}}, Piclo\footnote{\url{https://www.openutility.com/piclo/}}, Dajie\footnote{\url{https://www.dajie.eu/}}, Powerpeers\footnote{\url{https://www.powerpeers.nl/}}.

% P2P market designs and opt techniques
\newcommand{\sumP}{\ensuremath{\displaystyle\sum_{m\in\omega_n} P_{nm}}}

\section{Designs for peer-to-peer markets}\label{P2P_market}
Following \cite{Beitollahi2007, Parag2016} and the relevant literature, this section lists and discusses the P2P structures that have been proposed so far for P2P markets: (i) full P2P market; (ii) community-based market; and (iii) hybrid P2P market. The degree of decentralization and topology is what distinguishes them from each other and it can range from full P2P to hierarchical P2P structure.

\subsection{Full P2P market}\label{Full_P2P}
\begin{figure}[htb]
\centering
\includegraphics[width=8cm]{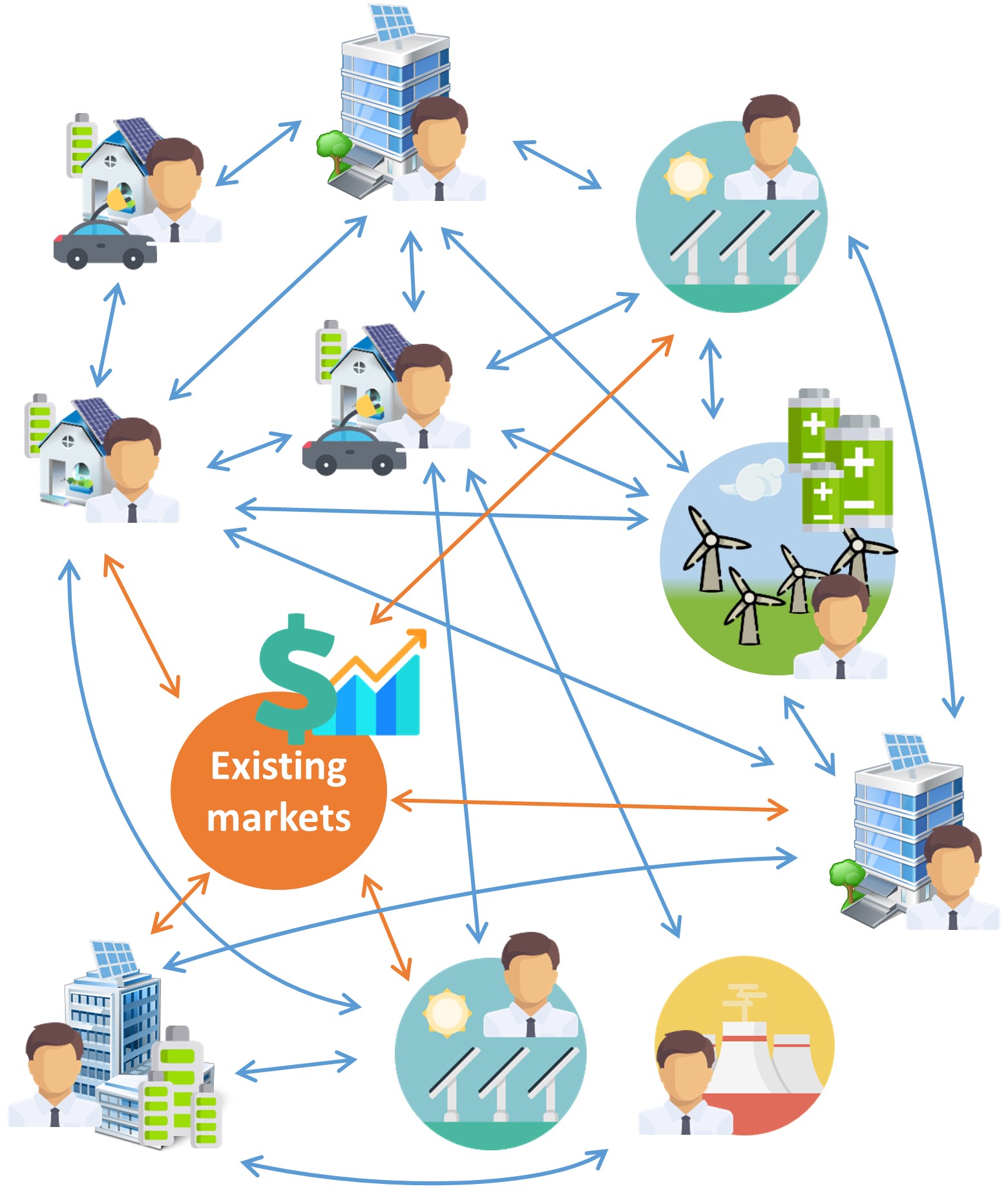}
\caption{Full P2P market market design.}
\label{Fig_Full_P2P}
\end{figure}

This market design is based on peers directly negotiating with each other, in order to sell and buy electric energy, as shown in Figure \ref{Fig_Full_P2P}.

Hence, two peers can agree on a transaction for a certain amount of energy and a price without centralized supervision. Sorin \textit{et al.} \cite{Sorin2017} proposed a full P2P market design between producers and consumers, which relies on a multi-bilateral economic dispatch. The P2P structure includes product differentiation where consumers can express their preferences, such as local or green energy. Morstyn \textit{et al.} \cite{Morstyn2018} implemented a P2P energy trading for real-time and forward markets of prosumers. Each agent's preferences that capture upstream-downstream energy balance and forward market uncertainty are included in the proposed framework.

\begin{figure}[htb]
\centering
\includegraphics[width=10cm]{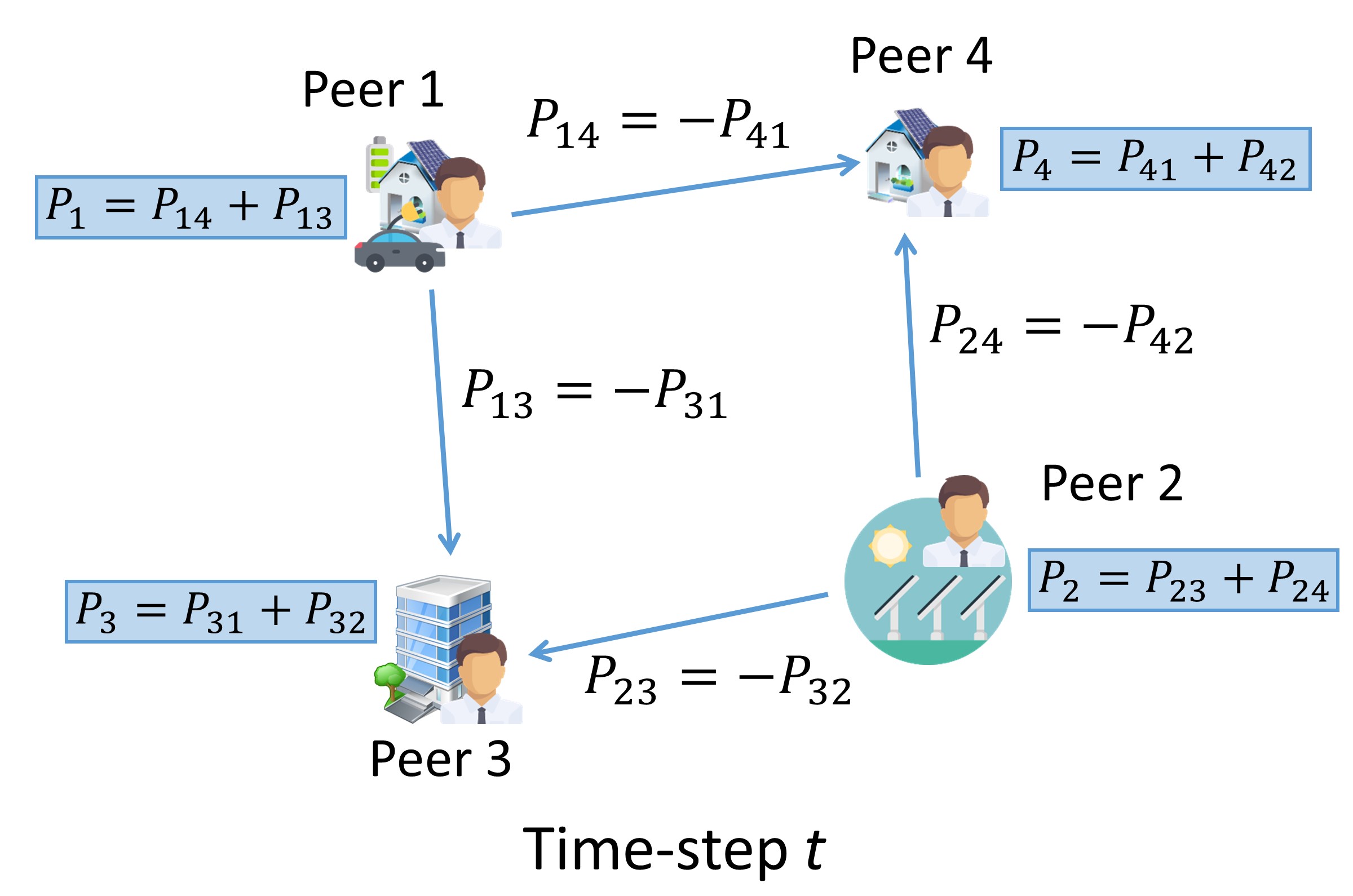}
\caption{Illustrative example of a full P2P market design.}
\label{Fig_Full_P2P_maths}
\end{figure}

In connection with the iconic Brooklyn experiment, a microgrid energy market is developed and published in \cite{Mengelkamp2017}. This framework enables a local microgrid market without central entity for small agents to trade energy locally. Alvaro-Hermana \textit{et al.} \cite{Hermana2016} implemented P2P energy trading between electric vehicles. The objective of the proposed approach is to increase bilateral trade between EVs, instead of them charging from the pool market. The recent research shows that this market design is gaining momentum in the industrial and academic fields. A general mathematical formulation of a full P2P market design is presented below, whereas more details can be found in \cite{Sorin2017}, which in its simplest form reads as
\begin{subequations}\label{eq:full_P2P}
\allowdisplaybreaks
\begin{align}
\label{eq:full_P2P_obj}
\min_{D}\quad&\sum_{n\in\Omega}C_n\left(\sumP\right)&&\\
\label{eq:full_P2P_trades}
\text{s.t.} \quad&\underline{P_n} \leq \sumP \leq  \overline{P_n} && \forall n \in \Omega\\
\label{eq:full_P2P_bal}
&P_{nm}+P_{mn}=0&&\forall (n,m) \in (\Omega,\omega_n) \\
\label{eq:full_P2P_prod}
&P_{nm}\geq 0 &&\forall (n,m) \in (\Omega_p,\omega_n) \\
\label{eq:full_P2P_cons}
&P_{nm}\leq 0 &&\forall (n,m) \in (\Omega_c,\omega_n)
\end{align}
\end{subequations}
where $D=(P_{nm}\in\mathbb{R})_{n\in\Omega, m\in\omega_n}$, with $P_{nm}$ corresponds to the trade between agents $n$ and $m$, for which a positive value means sale/production \eqref{eq:full_P2P_prod} and a negative value is equal to a purchase/consumption \eqref{eq:full_P2P_cons}. $\Omega$, $\Omega_p$ and $\Omega_c$ as sets for all peers, producers and consumers, respectively (hence, $\Omega_p, \Omega_c \in \Omega, \ \Omega_p \cap \Omega_c = \emptyset$). Figure \ref{Fig_Full_P2P_maths} shows a simple example to illustrate the P2P trading between 4 peers, in which peers 1 and 2 are producers, peers 3 and 4 are consumers. However, the above formulation can be readily generalized so that all peers are seen as prosumers, i.e., being able to both consumer and produce electric energy.

The set $\omega_n$ contains the trading partners\footnote{for consumers the producers and prosumers are trading partners, and {\it vice versa}} of a certain peer $n$. The bilateral trades $P_{nm}$ have reciprocity property, as defined by \eqref{eq:full_P2P_bal}, and for example, the power trades $P_{23}$ and $P_{32}$ have to be equal but with opposite signs. The associated dual variable $\lambda_{nm}$ represent the price for each bilateral trade. In principle, the outcome of the negotiation process can yield different prices for each and every trade. The function $C_n$ mostly corresponds to the production cost (or willingness to pay). A quadratic function \cite{Hug2015} is commonly used to represent production/consumption costs, using three positive parameters $a_n$, $b_n$ and $c_n$.

The optimization problem \eqref{eq:full_P2P} invites the use of decentralized or distributed optimization techniques \cite{Conejo2006, Boyd2010} in order to respect the basic nature of a P2P structure. Decomposition techniques such as Lagrangian relaxation, Alternating Direction Method of Multipliers (ADMM) and consensus+innovation are promising candidates. These make it possible to explicitly define individual problems for each agent, while guaranteeing their privacy. Each agent only shares the power and price that it is willing to trade, without revealing sensible information.

\subsection{Community-based market}\label{Community_market}
\begin{figure}[htb]
\centering
\includegraphics[width=8cm]{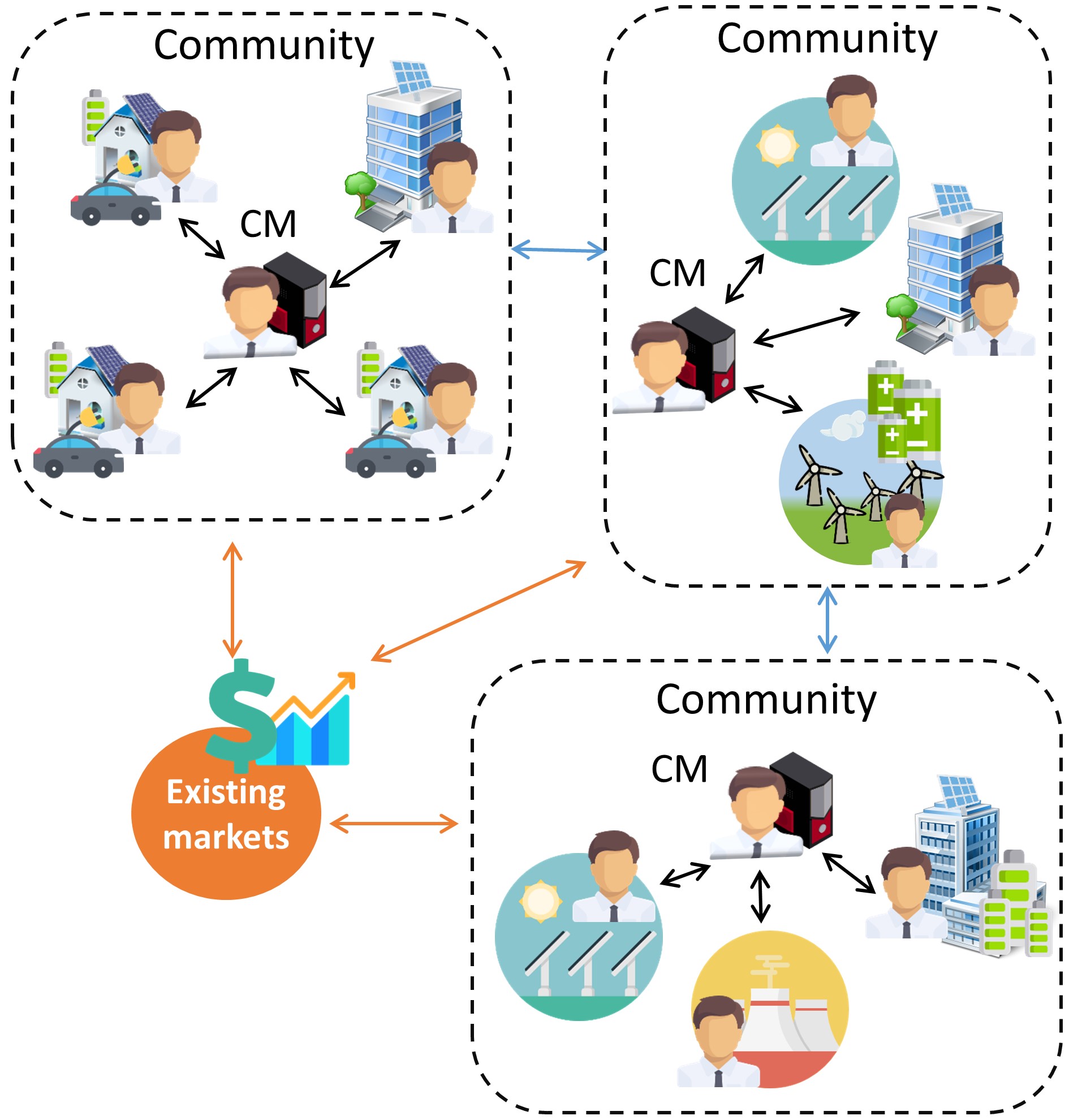}
\caption{Community-based market design.}
\label{Fig_Commu_Market}
\end{figure}

This design is more structured with a community manager who manages trading activities inside the community, as well as intermediator between the community and the rest of the system, as shown in Figure \ref{Fig_Commu_Market}.

\begin{figure}[htb]
\centering
\includegraphics[width=10cm]{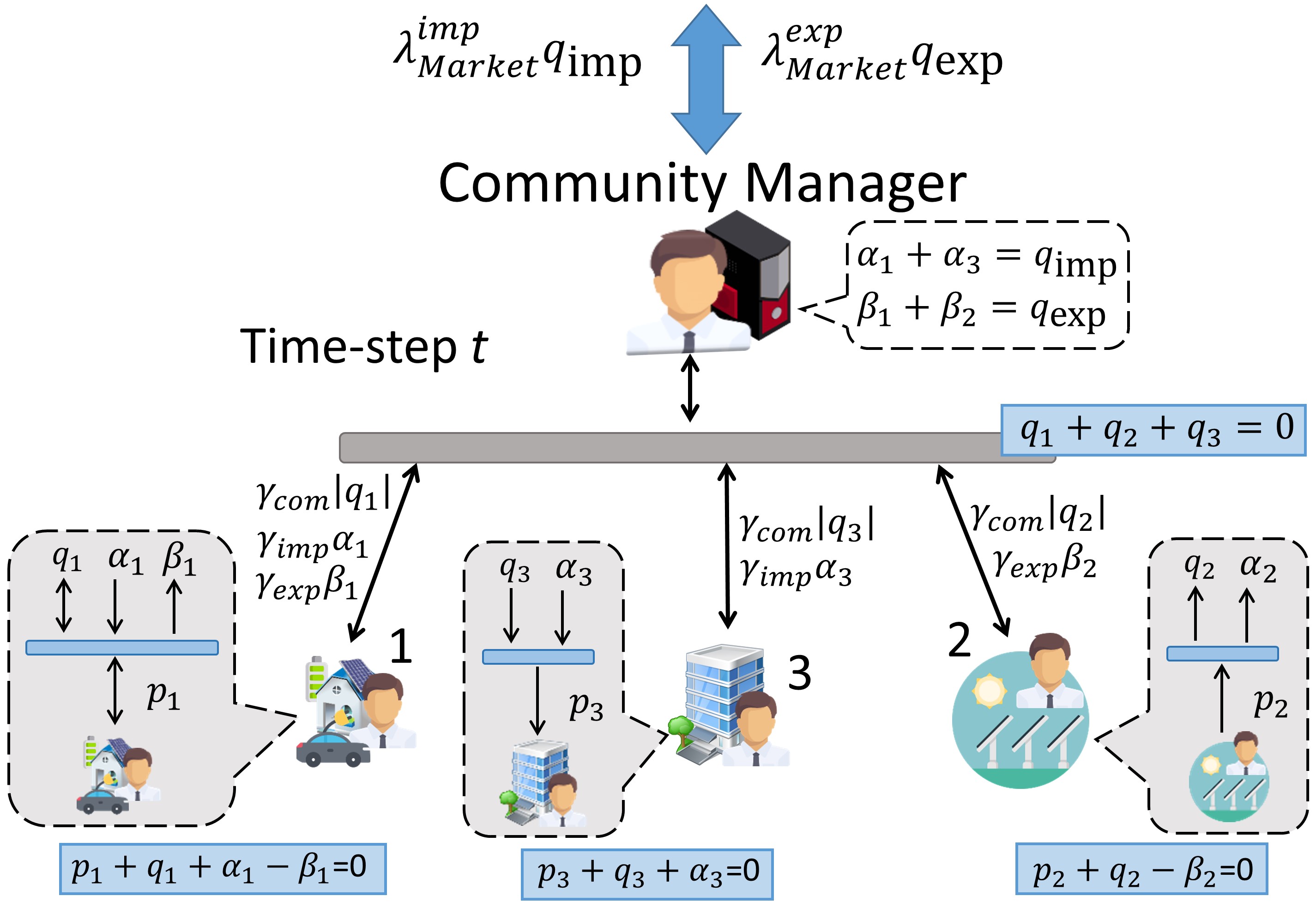}
\caption{Illustrative example of a community market structure.}
\label{Fig_Commu_Market_maths}
\end{figure}

This market design can readily be applied to microgrids \cite{Akter2016, Rosell2016} or to a group of neighbouring prosumers \cite{Verschae2016, Ilieva2016} that are natural constructs due to their location (i.e. being geographically close). More generally a community is to be based on members that share common interests and goals: for instance, a group of members that are willing to share green energy, though they are not at the same location. Moret and Pinson \cite{Moret2018a} formulated a community-based market with prosumers working in a collaborative manner. On the other hand, a multi-class energy management of a community-based market is designed by Morstyn and McCulloch \cite{Morstyn2018c}. The proposal formulated three different classes of energy to translate the prosumers' preferences. Tushar \textit{et al.} \cite{Tushar2016} implemented an auction scheme to share energy storage in a community. These are composed of agents with storage devices on the one hand, and on the other hand of agents willing to use those shared energy storage. A general mathematical formulation of a community-market design, following \cite{Moret2018a}, can be written as
\begin{subequations}\label{eq:community_market}
\allowdisplaybreaks
\begin{align}
\label{eq:comm_market_obj}
\min_{D} \quad& \sum_{n\in\Omega} C_n(p_n,q_n,\alpha_n, \beta_n)+G(q_{\text{imp}},q_{\text{exp}}) \\
\label{eq:comm_market_bal}
\text{s.t.} \quad& p_n + q_n + \alpha_n - \beta_n= 0 \; , \qquad \forall n \in \Omega \\
\label{eq:comm_market_trades}
& \sum_{n\in\Omega} q_n = 0 \\
\label{eq:comm_market_net_imp}
& \sum_{n\in\Omega} \alpha_n = q_{\text{imp}} \\
\label{eq:comm_market_net_exp}
& \sum_{n\in\Omega} \beta_n = q_{\text{exp}} \\
\label{eq:comm_market_prosumer}
& \underline{P_n}\leq P_n\leq \overline{P_n}\qquad \forall n \in \Omega \\
\end{align}
\end{subequations}
where $D=(p_n, q_n, \alpha_n, \beta_n \in\mathbb{R})_{n\in\Omega}$. $p_n$ corresponds to the production or consumption of peer $n$, depending on whether it is a producer or consumer, respectively. $\Omega$ is the set corresponding to all peers in the community. Figure \ref{Fig_Commu_Market_maths} shows a small community as illustrative example.

Each agent trades within the community through $q_n$ without knowing to which member, because it is centrally handled by the community manager through \eqref{eq:comm_market_trades}. Each peer can also choose to trade with the outside through $\alpha_n$ and $\beta_n$, which are respectively the power import and export. The sum of these trades is centrally handled by the community manager through \eqref{eq:comm_market_net_imp} and \eqref{eq:comm_market_net_exp}. The objective function \eqref{eq:comm_market_obj} accounts for the cost associated with all decision variables. Starting with a quadratic cost function of $p_n$, then a transaction cost $\gamma_{com}$ associated to $q_n$. For $\alpha_n$ and $\beta_n$, one can use weighting coefficients $\gamma_{imp}$ and $\gamma_{exp}$ translating the member's preference towards the outside world. The community manager also has a function associated to the energy exchanged with the outside world $G(q_{\text{imp}},q_{\text{exp}})$. This function can be modelled in different ways, but the most straightforward one readily links to day-ahead wholesale market prices.

As for the previous design, the negogiation process can be solved in a distributed manner \cite{Boyd2010}, for which there is a central node (community manager) to manage the remaining ones (agents). Each agent solves its own problem and only shares the required information with the central node. One can employ ADMM or similar distributed technique to solve the community-based problem in \eqref{eq:community_market}.

\subsection{Hybrid P2P market}\label{Hybrid_market}
\begin{figure}[htb]
\centering
\includegraphics[width=14cm]{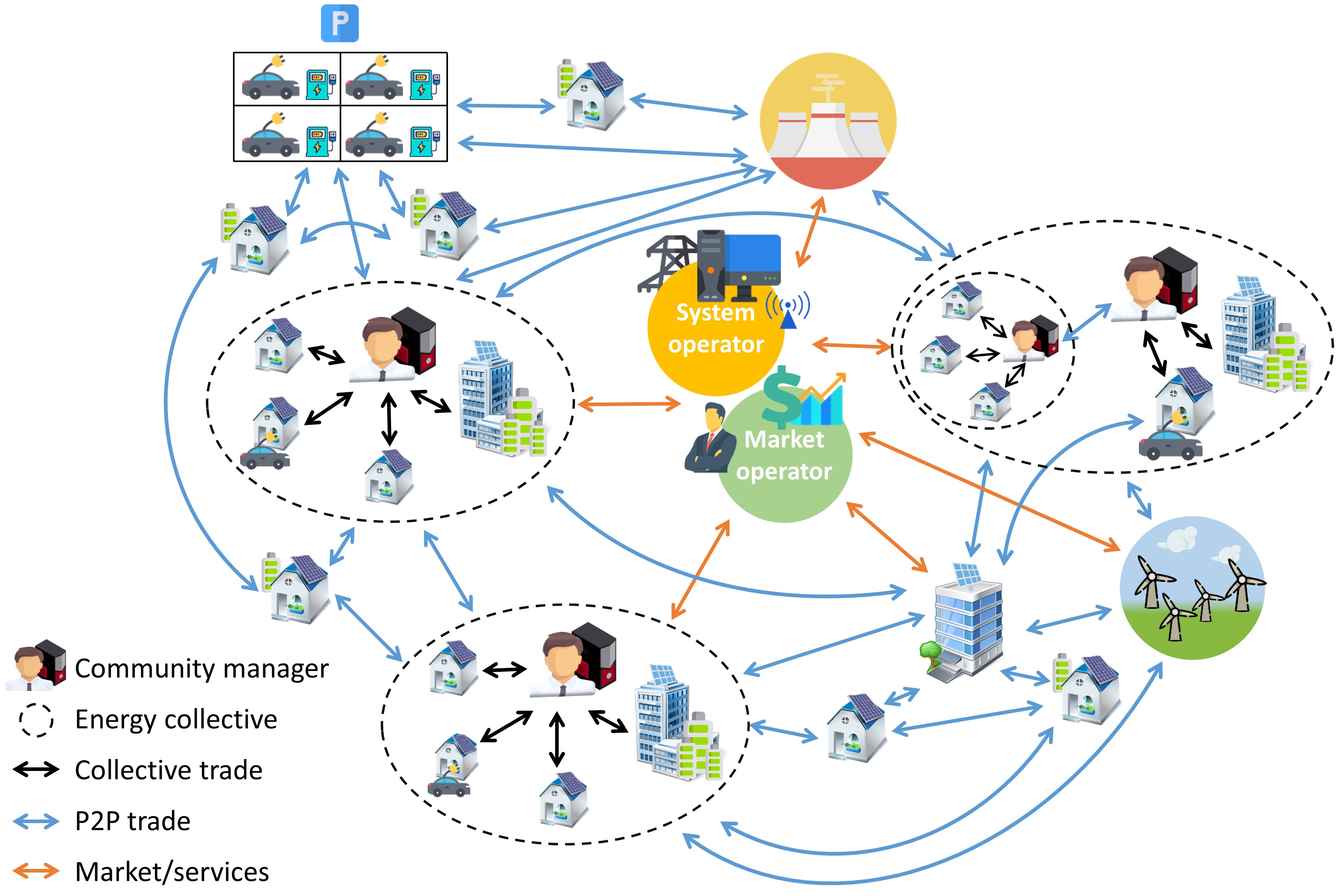}
\caption{Hybrid P2P market design.}
\label{Fig_Hybrid_P2P}
\end{figure}

This design is the combination of the two previous designs, ending up with different layers for trading energy, as shown in Figure \ref{Fig_Hybrid_P2P}. This proposal is seen as a ``Russian doll'' approach, where in each layer communities (or energy collectives) and single peers may interact directly with each other.

In the upper level, one finds individual peers or energy collectives engaging in P2P transactions between themselves, and also interacting with existing markets. In the bottom level, the energy collectives behave like the community-based approach previously introduced, where a community manager oversees the trading inside its community. As shown on the right of the picture, energy collectives can be nested into each other (e.g., buidlings and their inhabitants forming an energy collective, being part of another energy collective for the neighborhood). Although, there is no generic mathematical formulation to this hybrid P2P design, one can combine the two previous formulations to write a simplistic version of this design. Two levels are assumed in this formulation: (i) the bottom level only assumes communities using \eqref{eq:community_market}; (ii) the upper level assumes a P2P negotiation between individual peers and community managers using \eqref{eq:full_P2P}. The simplest form of this formulation reads as
\begin{subequations}\label{eq:hybrid_P2P}
\allowdisplaybreaks
\begin{align}
\label{eq:hybrid_P2P_obj}
\min_{D}\quad&\sum_{n\in\Omega^u}C^u_n\left(\sumP\right)+\sum_{n\in\Omega^b}C^b_n(p_n,q_n,\alpha_n, \beta_n)\\
\label{eq:upper_level}
\text{s.t.} \quad&\text{upper level - full P2P design:} \nonumber \\
& \quad \text{constraints in \eqref{eq:full_P2P}} && \forall n \in \Omega^u\\
\label{eq:commu_P2P_bal}
& \quad \sumP=q^n_{\text{exp}}-q^n_{\text{imp}} && \forall (n,m) \in (\Omega_{co},\omega_n) \\
\label{eq:bottom_level}
&\text{bottom level - community-based design:} \nonumber \\
& \quad \text{constraints in \eqref{eq:community_market}} && \forall n \in \Omega^b
\end{align}
\end{subequations}
where $D=(P_{nm}\in\mathbb{R}_{n\in\Omega^u}, p_n, q_n, \alpha_n, \beta_n \in\mathbb{R}_{n\in\Omega^b})$. $\Omega^u$ and $\Omega^b$ are sets for all peers in the upper and bottom level, respectively (hence, $\Omega^u \cap \ \Omega^b = \Omega$). In the bottom level, the community manager of each community $n \in \Omega^b$ determines the internal energy needs\footnote{community internal consumption or production)} $q_n$ plus the desire energy import ($q^n_{\text{exp}}$) or export ($q^n_{\text{imp}}$). Then, the full P2P market formulation is used in the upper level to calculate the optimal P2P energy trading between the peers $n \in \Omega^u$ (i.e. individual prosumers and community managers). The sum of bilateral trades $\sum\nolimits_{m\in\omega_n}P_{nm}$ is equal to the amount of $q^n_{\text{exp}}$ minus $q^n_{\text{imp}}$ defined by the community manager of each community $n$ \eqref{eq:commu_P2P_bal}.

Some authors started to explore this nested approach, such as Long \textit{et al.} \cite{Long2017} with a hybrid design containing three distinct levels for a distribution grid. The upper level assumes the grid divided into cells trading among themselves. In the second level, trades happen between microgrids under the same cell. At the lower level, a community-market design is applied for each microgrid. In \cite{Liu2015}, an hybrid approach is proposed for microgrids under the same distribution grid, where the grid constraints are included in the P2P trading between microgrids. This work uses a relaxed formulation of an AC optimal power flow and it removes the price and negotiation mechanism between microgrids.

\subsection{Comparison on P2P market designs}
The literature converges so far on three different market designs for P2P markets, even if some references use different terms to describe the same type of market structure. The main advantages, challenges and references of the three P2P market designs are presented in Table \ref{P2P_market_description}.

\begin{table}[]
\centering
\caption{Summary of three P2P market designs (based on \cite{Parag2016}).}
\label{P2P_market_description}
\resizebox{\textwidth}{!}{%
\begin{tabular}{llll}
\hline
P2P market structure & Main advantages & Main challenges & References \\ \hline
Full P2P market & \begin{tabular}[c]{@{}l@{}}1) Total freedom of \\ choice and autonomy,\\ empowering the active \\ consumers;\\ \\ 2) Energy use aligned \\ with each agent's \\ preference (e.g. cost, \\ green, local, etc);\\ \\ 3) Complete "democratization" \\ of energy use\end{tabular} & \begin{tabular}[c]{@{}l@{}}1) Investment and maintenance \\ with ICT infrastructure \\ in case of scalability\\ to all system;\\ \\ 2) Potential slow convergence \\ to obtain a consensus in the \\ final delivery of energy;\\ \\ 3) Predicting system behaviour \\ by grid operators, because of \\ the lack of centralized control;\\ \\ 4) Guarantee of safety \\ and high-quality energy \\ delivery\end{tabular} & \begin{tabular}[c]{@{}l@{}}\cite{Mengelkamp2017}, \cite{Sorin2017},\\ \cite{Sorin2017b}, \cite{Morstyn2018},\\ \cite{Hermana2016}\end{tabular} \\ \hline
\begin{tabular}[c]{@{}l@{}}Community-based \\ market\end{tabular} & \begin{tabular}[c]{@{}l@{}}1) Enhancing the relationship \\ and involvement of \\ community members,\\ because of sharing a \\ common good (i.e. energy);\\ \\ 2) Mobilizing social \\ cooperation and resilience\\ in community members;\\ \\ 3) Potential new services \\ for grid operators provided \\ by the community manager\end{tabular} & \begin{tabular}[c]{@{}l@{}}1) Reaching the preferences \\ of energy use for all \\ community members at all time;\\ \\ 2) For the community manager\\ is aggregating all members' data \\ and managing their expectations;\\ \\ 3) Having a fair and unbias \\ energy sharing among \\ community members\end{tabular} & \begin{tabular}[c]{@{}l@{}}\cite{Liu2017}, \cite{Ilic2012},\\ \cite{Akter2016}, \cite{Verschae2016},\\ \cite{Rosell2016}, \cite{Ilieva2016},\\ \cite{Kang2017}, \cite{Moret2018a},\\ \cite{Morstyn2018c}, \cite{Tushar2016}\end{tabular} \\ \hline
Hybrid P2P market & \begin{tabular}[c]{@{}l@{}}1) ICT infrastructure and \\ computation effort are \\ scalable to all system;\\ \\ 2) Most compatible with \\ the system in the next \\ years, it can be seen\\ as co-existent design \\ of the two previous ones;\\ \\ 3) More predictable to \\ the grid operators\end{tabular} & \begin{tabular}[c]{@{}l@{}}1) Coordinating internal \\ trades in the communities \\ with trades between\\ high-level agents (e.g. \\ community managers, \\ utilities, etc)\end{tabular} & \cite{Long2017}, \cite{Liu2015} \\ \hline
\end{tabular}%
}
\end{table}

% Pros and cons of full P2P market
The first design gives the possibility that the energy usage is truly aligned with consumers' preferences. On the other hand, there is a scalability problem concerning the negotiation process, as analyzed by the authors in \cite{Moret2018b}. The negotiation can become a computational burden for scenarios with many participants in this design. If this design is used for the entire power system, the scalability is a real challenge yet to be solved. One way to handle this issue could be through sparse graphs to reduce the number of communications.

% Pros and cons of Community market
On the other hand, the main advanges of the community-based market design are the enhancement of involvement and cooperation of community members. The fact of being a more structured design allows the community manager to provide services to the grid operators as an aggregator. The revenues from these services can be shared (e.g., in a proportional way) by all community members. Even if all members agree to participate in a community, there may be times when a member's expectation is not aligned with the community's general interests.

%Compared to the full P2P considered previously, there is more research and published literature on this type of market design. In addition, one can find a sustantial amount of work concerning control architectures to organize communities, overlooking market considerations, such as microgrid control and aggregator-based operation \cite{Nosratabadi2017, Burger2017}.

% Pros and cons of Hybrid P2P market
The final design shares some of the advantages of the two previous designs such as empowering choice of agents, increasing their involvement and cooperation. Besides that, the scalability of all the system is in a way covered by this design, because the ICT and computational effort required are less than required by the full P2P market design. The two previous market designs can co-exist and interact in such a hybrid structure, where the coordination of trades at and between levels is important.

% ICT and blockchain
Another important aspect to mention is related the ICT infrastructure that must sustain these market designs. The proposals explored have in common two main pillars, i.e the physical and virtual layers. In most of the cases, the protocols proposed for smart grid applications are used in the physical layer \cite{Jogunola2017}, while blockchain technology\footnote{distributed platform for data management that can register and settle transactions between peers} \cite{Sikorski2017, Drescher2017} has received attention as potential solution for the virtual layer. This technology is the backbone behind the crypto-currencies that have appeared in recent years\footnote{bitcoin is the most famous case}. Some literature argue that blockchain can be the key factor to deploy a P2P market in the energy sector \cite{PWC2016, Andoni2017, Vangulick2018}. There are enormous advantages such as data management without third-party supervision, but caveats such as scalability and data storage need to be addressed before a real implementation in the energy sector \cite{Bozic2016}. Although, many consider blockchain the most important asset in a successful deployment of P2P markets, it is worth mentioning that such markets can exist without blockchain.

% Opportunities and challenges
\section{Opportunities and challenges}\label{Opportunities_challenges}
This section focuses on future prospects with P2P markets, by starting with an analysis of opportunities and remaining challenges. Then, topics worthy of investigation by the research and industrial communities in the coming years are introduced and discussed.

\subsection{Discussion on P2P market potential}\label{SWOT_analysis}
For a better analysis of potential opportunities and challenges about this topic, the authors elaborated an analysis based on Strengths, Weaknesses, Opportunities and Threats (in short SWOT) that is shown in Table \ref{tb_Opp_Chall}. The main enablers and obstacles supported our analysis concerning P2P market potential.

\begin{table}[htb]
\centering
\caption{Summary of potential strengths, weaknesses, opportunities and threats}
\label{tb_Opp_Chall}
\resizebox{\textwidth}{!}{%
\begin{tabular}{llll}
\hline
Strengths & Weaknesses & Opportunities & Threats \\ \hline
\begin{tabular}[c]{@{}l@{}}1) Empowerment of \\ consumers, focusing\\ in trust, transparency\\ and openness\end{tabular} & \begin{tabular}[c]{@{}l@{}}1) Sub-optimal energy \\ price of all energy \\ system\end{tabular} & \begin{tabular}[c]{@{}l@{}}1) Democratization\\   of energy\end{tabular} & \begin{tabular}[c]{@{}l@{}}1) Legal and regulatory \\ obstacles, which \\ influence the transition\\ to these markets\end{tabular} \\
\begin{tabular}[c]{@{}l@{}}2) Consumers have better \\ choice of supply and\\ possibility to  produce \\ and sell their own energy\end{tabular} & \begin{tabular}[c]{@{}l@{}}2) Potentially overwhelming\\ transition to this \\ consumer-centric market\end{tabular} & \begin{tabular}[c]{@{}l@{}}2) Increase consumers \\ awareness and cooperation \\ towards environmental \\ energy consumption\end{tabular} & \begin{tabular}[c]{@{}l@{}}2) Energy poverty for\\ some group of \\ consumers\end{tabular} \\
\begin{tabular}[c]{@{}l@{}}3) Increase resilience and \\ reliability of the system\end{tabular} & \begin{tabular}[c]{@{}l@{}}3) Heaviness of negotiation\\ and clearing mechanisms\end{tabular} & \begin{tabular}[c]{@{}l@{}}3) Create new \\ business models\end{tabular} & \begin{tabular}[c]{@{}l@{}}3) Prosumer engagement\\ and its human dimension\end{tabular} \\
\begin{tabular}[c]{@{}l@{}}4) Remove potential market\\ power from some players\\ in the wholesale market\end{tabular} & \begin{tabular}[c]{@{}l@{}}4) Life-cycle assessment\\ of hardware infrastructure\end{tabular} & \begin{tabular}[c]{@{}l@{}}4) Boost retailer market, \\ since lacks competition\end{tabular} & \begin{tabular}[c]{@{}l@{}}4) Potential grid \\ congestions\end{tabular} \\
 &  & \begin{tabular}[c]{@{}l@{}}5) Postpone grid investments \\ from system operators\end{tabular} & \begin{tabular}[c]{@{}l@{}}5) Technology \\ dependency\\ (e.g. blockchain)\end{tabular} \\
 &  &  & \begin{tabular}[c]{@{}l@{}}6) Security and privacy\\ with data\end{tabular} \\
 &  &  & \begin{tabular}[c]{@{}l@{}}7) Potential failure of\\ these markets if \\ poorly structured\end{tabular} \\ \hline
\end{tabular}%
}
\end{table}

The empowerment of consumers choice and transparency is an obvious strength of P2P markets, because of their flexibility that enables consumers choice on the type and origin of their electricity in a dynamic manner. Thus, consumers create empathy towards those who collaborate, particularly when they receive positive feedback. Burdermann \textit{et al.} \cite{Brudermann2014} indicated five different classes of collaboration among consumers in P2P markets for an urban area. The system resilience and security can improve, because consumers will be compelled to solve problems and not jeopardize the normal operation of others, especially the ones they know from previous transactions. We can have cases like unexpected loss of renewable production or congestion problems in the grid that can be solved in a collaborative manner by all involved peers (i.e. large producers and small prosumers). This may lead to a rethink of the top-down hierarchical structure used for grid operation. Finally, prosumers will be less volatile to the wholesale market price, because they found alternative solutions to their energy supplier by engaging in P2P trading/sharing approaches.

The P2P design allows product differentiation reflecting consumers' preferences in energy trading that leads to different prices for every transaction. However, this aspect can also lead to a sub-optimization of the overall energy price, which represents an obvious weakness in these markets. The fact that all market participants simultaneously negotiate with all others can help prices to converge to similar values. Further investigation is required to determine the magnitude of this sub-optimal price. The overwhelming number of transactions and the heaviness of the negotiation mechanism are other weaknesses as investigated in \cite{Moret2018b}. The first one is particularly important for a full P2P design, but it can be mitigated in other less "anarchical" designs, like community-based or hybrid P2P. The life-cycle assessment of certain hardware supporting the negotiation process (e.g. smart devices, batteries, PV panels, etc) can be a weakness in such P2P markets, which future investigation has to pay attention to. The economy of scale in this P2P markets approach can be one way to mitigate this weakness.

This new form of market can create opportunities for all participants in the electric power system. The first opportunity is more democratization of energy by adopting such a market approach. The introduction of such new markets also creates the possibility of having new business models. Currently, the retail market lacks competition in many countries, for which the P2P market can be a solution. Retail companies may have to adapt their business to more P2P transactions. Finally, the grid operators can benefit from such markets by deferring grid investments in new lines and equipment. One of the strengths is the increase in resilience and security, which creates an opportunity to solve grid problems by all market participants rather than reinforcing the grid.

If we look to the threats associated with P2P markets, the main threat concerns the legal framework in most countries. However, the authors believe that with time this obstacle will be less and less important and eventually removed. Energy poverty of some consumers or communities with less economic power can arise when P2P markets are implemented. In addition, consumer and prosumer engagement, i.e. their willingness to participate and fully utilize the possibilities offered by such a decentralized, is also a point of concern since interest in electricity matters is commonly low. Different studies \cite{Ploug2017, Pallesen2018, Gyamfi2013, Allcott2010} emphasise how important it is to account for human behaviour in electricity markets. The EU is also committed, with several projects, to bridge the gap between technology-based potential of consumer programmes and actual market behaviour. More precisely, this was the central focus of projects such as BEHAVE\footnote{\url{https://ec.europa.eu/energy/intelligent/projects/en/projects/behave}}, Penny\footnote{\url{http://www.penny-project.eu/}} and PEAKapp\footnote{\url{http://www.peakapp.eu/}}. P2P market development ought to account for the insights gained through this type of projects, to eventually reduce the potential of this threat. Bounded rationality of electricity consumers and prosumers may also play a role in market design and operation \cite{Simon1997}. Finally, academic and industrial partners that work on this topic must be aware that a poor design can cause a potential failure in P2P markets. Poorly designed markets could indeed have an effect on system resilience and power system security that is the opposite of what P2P systems are normally praised for. The exploitation of the positive aspects (strengths) and resolution of the handicaps (weaknesses) would result in successful implementation of P2P markets in its different structures. The next sub-sections discuss new directions based on this analysis that are worth exploring in the coming years.

\subsection{Recommendations on business models and grid operation}\label{Business_and_grid}
% Recommendation on Business models
So far, the paper has explored the implementation of P2P markets as alternative structures to the existing ones in the electricity market, i.e. wholesale and retail. On top of this layer, the electricity market has different type of business models operated by different actors (e.g. utilities, retailers, etc). The business models can be classified as C2C, B2C and B2B based on how different kind of market actors trade good or services, where C2C, B2C and B2B stands for Consumer-to-Consumer, Business-to-Consumer and Business-to-Business, respectively. For the electricity market, a retailer with a business model that intermediates the energy trading between an electric utility and small consumer is classified as B2C business model. This can be operated in one of the structures defined for the electricity market (e.g. wholesale, retail or P2P). Today, the existing B2C business models heavily rely on the centralized pool-based structure used in the wholesale market, thereby prosumers are unaware, and cannot select directly which utilities real provide their energy. On the other hand, there are companies (e.g. Vandeborn and Piclo), as shown in Section \ref{P2P_status}, using P2P designs that offer as business the intermediation of the energy trade between small consumers and local renewable producers, whereas it is within a C2C business context. This has a residual effect in the entire electric power system, because of the small scale P2P application and lack of direct connection through B2C models. Therefore, researchers has been challenging decision-makers to see P2P as a new channel for B2C business models that could allow large utilities to directly trade energy with small prosumers. This could be done through direct mechanisms for real-time negotiation and endogenous consideration of preferences, instead of the statistical accounting performed today for the case of guarantees of origin. However, new B2C business models have, in its essence, to follow the consumer-centric premise that characterizes P2P markets. Thus, it would create a sense that everyone is part of (and benefits from) this energy revolution towards a more sustainable electricity sector.

% Recommendation on grid operation
Grid operation under a P2P market is still a concern, mainly to grid operators. There are only few studies assessing the impact of P2P trade on grid operation \cite{Baroche2018, Guerrero2018}. In fact, grid congestion can be a potential threat if not properly handled, as indicated in Table \ref{tb_Opp_Chall}. However, P2P markets also present a new opportunity to rethink the use of common grid infrastructures and services, because P2P structures may allow the mapping of the energy exchanges. For example, the grid cost may depend on electrical distance associated to each P2P transaction. The recent breakthroughs concerning distributed optimization on grid operation \cite{Kargarian2017} can also inspire other works on the redesign of grid operation under P2P markets. Rethinking grid operations can deploy new business models, where communities or individual agents participate in flexibility services to respect the grid operation. This represents another opportunity to mobilize customers' flexibility and resilience through increased awareness and involvement.

% Test example
\section{Reference test case for P2P markets}\label{Test_case}
\begin{figure}[htb]
\centering
\includegraphics[width=12cm]{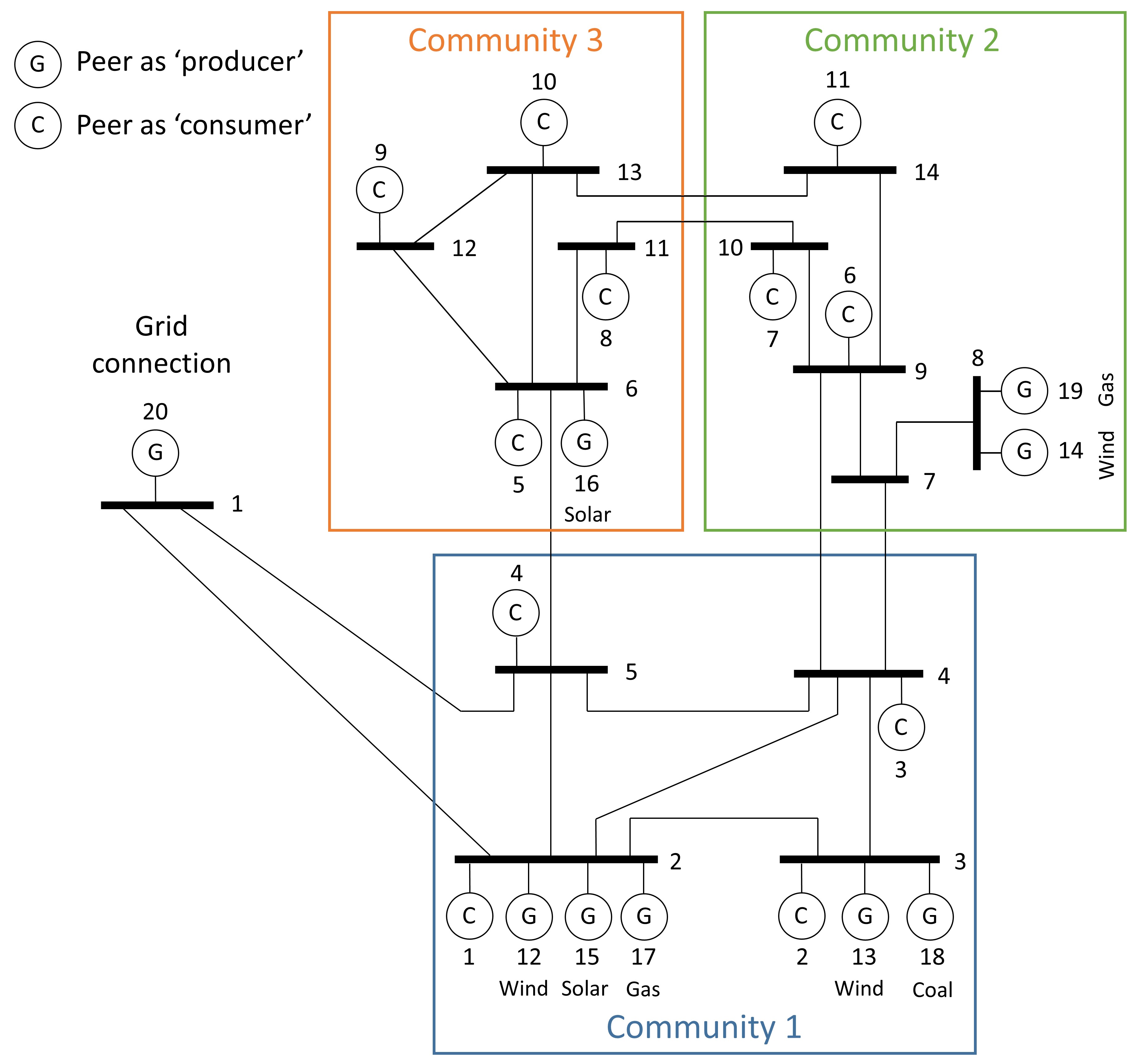}
\caption{IEEE 14-bus network system.}
\label{fig_IEEE_14_bus}
\end{figure}

In this section, the three P2P market designs described in Section \ref{P2P_market} are evaluated on the IEEE 14-bus network system presented in \cite{IEEE_14_bus}. This test case provides a realistic case to simulate P2P market designs, since there does not exist today a reference test case that would encourage reproducibility and benchmarking. The bus 1 is the upstream connection with the main grid, where the generator assumes an infinite power\footnote{lower and upper bound $\underline{P_n}$ and $\overline{P_n}$ are equal to $-\infty$ and $+\infty$, respectively}. Figure \ref{fig_IEEE_14_bus} shows the IEEE 14-bus system divided into three communities, containing 19 peers represented by their ID and type\footnote{(G) - producer, and (C) - consumer}.

The test case is simulated over one year with 30 minutes time-step based on available Australian data. More precisely, the production levels of wind turbines and PV plants, as well as the consumption of households, are taken from \cite{Dowell2016, Ratnam2017}. These data sets have been normalized and then scaled to the capacity of the wind turbines, PV plants and loads. A quadratic cost function is used to characterize the cost function of all resources, however the PV and the wind turbines are modeled as must-take producers ($\underline{P_n}=\overline{P_n}$ and $a_n=b_n=0$). The linear cost $b_n$ of peer 20 in bus 1 is equal to the market price from July 2012 to June 2013 provided by the Australian Energy Market Operator. This test case assumes a tariff of 10 \$/MWh for using the main grid. Thus, the importation price from the producer in bus 1 is equal to the market price plus the grid tariff. On the other hand, the export price is equal to the market price minus the grid tariff. Readers interested in using the data presented in the test case for further research are directed to \cite{P2P_test_case2018}.

The three market designs presented in Section \ref{P2P_market} are simulated using the mathematical formulation  \eqref{eq:full_P2P}, \eqref{eq:community_market} and \eqref{eq:hybrid_P2P}. The three communities defined in the test case are only used for the community and hybrid P2P designs. In the former, each community only trades with the main grid. In the latter, the three communities trade with each other and the main grid. Besides the quadratic cost function, the cost of $\gamma_{com}=0.001 \text{\$/MWh}$ is applied to all $P_{nm}$ in the full P2P design, as a transaction cost. For community and hybrid P2P designs, the same transaction cost is assumed for the energy traded ($q_n$) within each community \eqref{eq:comm_market_trades}. Besides that, the hybrid P2P design has transaction costs for the P2P trades between the three communities:
\begin{itemize}
\item Community 1-2: 2 \$/MWh
\item Community 1-3: 1 \$/MWh
\item Community 2-3: 1.5 \$/MWh
\end{itemize}

Each design is then solved by a centralized optimization approach. Table \ref{tb_P2P_IEEE_14_res} shows the results of all simulations, namely the social welfare, total import cost and export revenue to the main grid.

\begin{table}[htb]
\centering
\caption{Results of all P2P market designs.}
\label{tb_P2P_IEEE_14_res}
\resizebox{\textwidth}{!}{%
\begin{tabular}{llll}
\hline
Market designs & Total SW {[}M\${]} & Import cost {[}M\${]} & Export revenue {[}M\${]} \\ \hline
Full P2P & 45.21 & 0.072 & 56.66 \\
Community & 44.27 & 2.88 & 58.95 \\
Hybrid P2P & 44.32 & 2.86 & 58.71 \\ \hline
\end{tabular}%
}
\end{table}

The highest social welfare is reached with the full P2P design, while the community design presents the lowest result with a difference around 2\%. This is true, since the communities only trade with the main grid, while in the full P2P design it is possible to share the renewable surplus among all peers. In this test case, the exportation to the main grid is less profitable than trading with other peers between the three communities. Therefore, the hybrid P2P design improves the social welfare when compared to the community design. This improvement is affected by the extra transaction costs of 17.4k\$ between communities. When this cost is removed, the social welfare of the hybrid P2P design is similar to the full P2P design. The energy exchange in all P2P market designs is shown in Table \ref{tb_P2P_IEEE_14_res_2}, namely the total load, import, export and energy trade between communities.

\begin{table}[htb]
\centering
\caption{Energy trade for all P2P market designs.}
\label{tb_P2P_IEEE_14_res_2}
\resizebox{\textwidth}{!}{%
\begin{tabular}{lllll}
\hline
Market designs & \begin{tabular}[c]{@{}l@{}}Total \\ load {[}GWh{]}\end{tabular} & \begin{tabular}[c]{@{}l@{}}Total \\ import {[}GWh{]}\end{tabular} & \begin{tabular}[c]{@{}l@{}}Total \\ export {[}GWh{]}\end{tabular} & \begin{tabular}[c]{@{}l@{}}Community \\ exchange {[}GWh{]}\end{tabular} \\ \hline
Full P2P & 401.4 & 2.1 & 1041.1 & 54.4 \\
Community & 395.1 & 45.4 & 1093.4 & 0 \\
Hybrid P2P & 395.7 & 44.7 & 1085.5 & 9.1 \\ \hline
\end{tabular}%
}
\end{table}

\begin{figure}[htb]
\centering
\includegraphics[width=13cm]{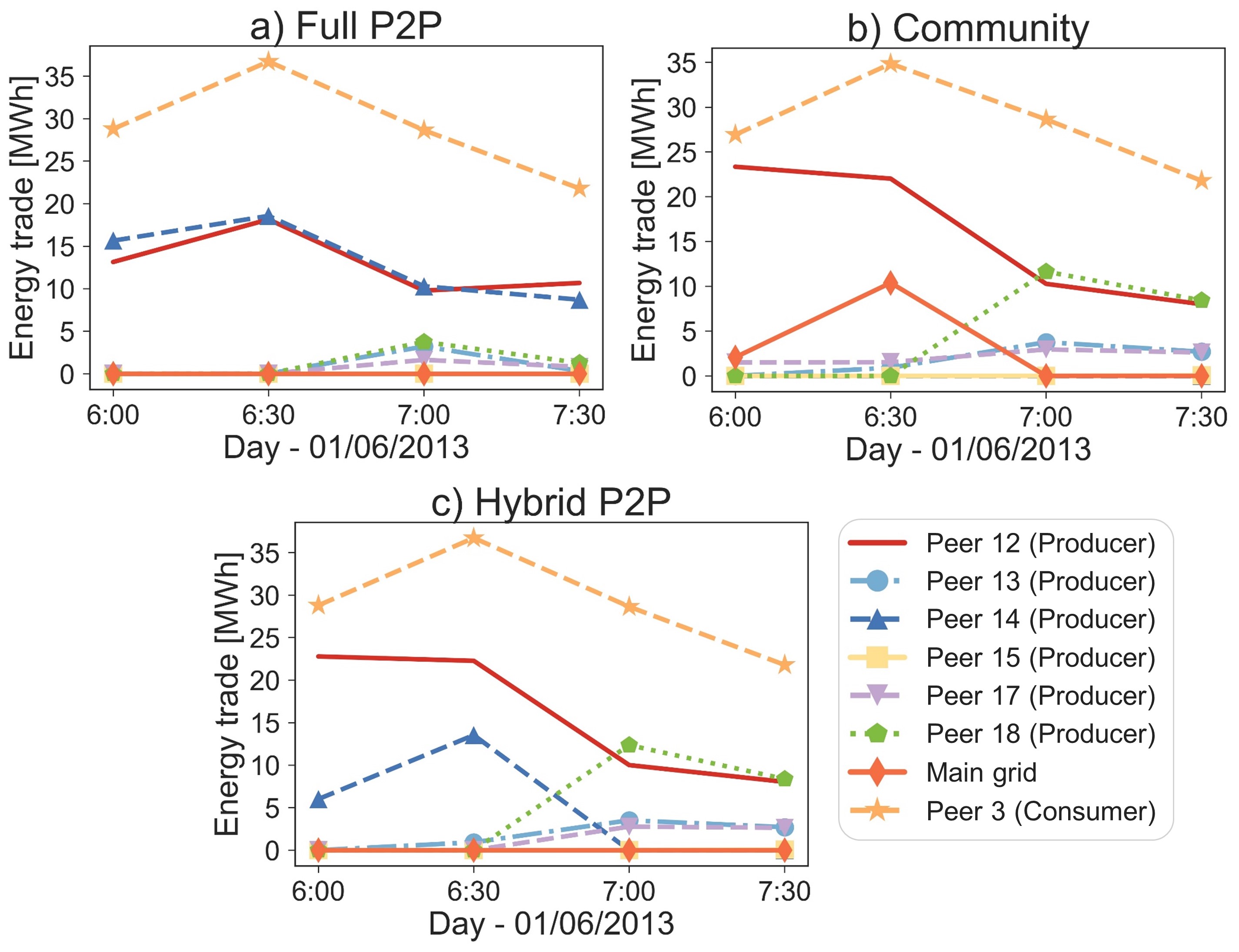}
\caption{Trades negotiated by consumer 3 at day 01/06/2013 from time step 6:00 to 7:30: a) full P2P, b) community and c) hybrid P2P.}
\label{fig_market_results}
\end{figure}

The full P2P design achieved the highest consumption, due to the same reason (sharing renewable surplus) mentioned before in Table \ref{tb_P2P_IEEE_14_res}. This fact influences the low energy import and high energy exchange between communities achieved in this design. On the other hand, the community design presents the highest results for the energy import and export. When both designs are compared, the energy import and export are reduced by 95\% and 5\%, respectively. The energy exchange between communities in the full P2P design (54.4 GWh) is almost the same as the energy export decrease (52.3 GWh). To illustrate the differences in trade among peers between the three P2P designs, the trade negotiated by peer 3 at day 01/06/2013 from time step 6:00 to 7:30 is shown in Figure \ref{fig_market_results}. The peer 3 belongs to the community 1 and its total consumption is represented by a yellow line, which also corresponds to the sum of all other lines.

The main difference between the three designs is the energy trade with peer 14 located in community 2 (represented by a blue line). For the full P2P design (Figure \ref{fig_market_results} a)), peer 14 is one of the top agents that peer 3 trades with during these time steps, namely at 6:00 and 6:30. In the hybrid P2P design (Figure \ref{fig_market_results} c)), the amount of trade with peer 14 is reduced because of the transaction cost of 2 \$/MWh for the energy trade between community 1 and 2. To avoid this extra cost, peer 3 trades more energy with the peers of the same community 1 (i.e. peers 12, 13, 15, 17 and 18). Similar behavior can be seen in the results of the community design (Figure \ref{fig_market_results} b)). In conclusion, the best result is obtained by the full P2P design when all peers share the renewable surplus. However, one should not conclude that this is the best design, since the main objective is to have a real test case for assessing different P2P designs. Moreover, the results may change when the network constraints are included in these market designs. Other researchers are free to use this test case in their investigations such as to validate new market designs and business models, or to assess new grid operation strategies.
% Conclusion
\section{Conclusions and perspectives}\label{Conc_recom}
% New conclusion
The current paradigm has the drivers to conduct the transition towards P2P markets in order to bring prosumers into power system operational practice. This paper contributes to such a discussion through a detailed review of P2P market proposals, as well as pointing at future areas of relevant research. Furthermore, it is concluded that there are conditions to deploy P2P markets in co-existence with existing market structures, as long as potential conflicts with historical actors are prevented since these are key to a smooth and manageable transition towards P2P markets. Future research should promote ways for P2P markets to be coupled with the existing wholesale and retail markets, allowing consumers to switch from one market to the other when it is most convenient.

Moreover, this study shows that P2P market may give a new taste to B2C business models for electricity, most likely with greater regard for consumer preferences and interests. Further contribution should be promoted on how to quantify the benefits and impacts of new forms of B2C models. One should also pay attention to how system operators could be involved in P2P markets, and satisfying them in terms of feasibility, reliability and security of supply. Most importantly, how to maintain high levels of power system reliability through the distributed provision of reserves. For example, probabilistic matching and queueing theory to solve imbalances of different prosumers in a collaborative manner is recommended by the authors.

From this literature review, one concludes that the hybrid P2P market design is the most suitable in terms of scalability, giving room for all other P2P designs to interact. Besides that, P2P designs invite the use of distributed optimization techniques that respect the privacy of every peer, and future work should aim to improve the negotiation processes. Arguably, scalability when reaching large number of peers is a current challenge. Sparsification of communication and negotiation graphs will be fundamental to reduce exchanges among peers with residual effect on the optimality of resource allocation and pricing outcomes. The research of methods to handle asynchronous communication is also a relevant future work. The last relevant area of future research should target the human dimension modelling of consumers, such as bounded rationality and strategic behaviour, which may drift towards the social science side to find ways to optimally express consumer preferences and appraise the impact of such preferences on market functioning and outcomes.

\appendix
\section{Review methodology}\label{research_method}
This review methodology follows the procedures outlined in \cite{Webster2002, Brocke2009}. The authors started by gathering technical reports, scientific papers and books from different fields, since our original goal was to invest in a broad and interdisciplinary review on P2P electricity markets. This included literature on energy, power systems, economics, operational research, computer and social sciences. At this stage, only peer-reviewed journal articles, books and conference proceedings in English were analysed in this review. This search led to a total of 112 publications. The next step was the analysis of each document to evaluate its relevance to the topic. Categories related to the research topic were defined\footnote{e.g. premises to P2P markets, projects or companies, P2P market designs, etc} and every publication was associated with at least one category. Three labels classified the literature to quantify their relevance in each category: A for a publication with high quality, B for a publication with some relevance to the category but important to the context of the review, whereas C for limited relevance, which led to exclude them. 

After this process, a total of 80 publications remained that were of high relevance, where there are 69 references with label A plus 11 references with label B. Focusing in references with label A, while 15 publications related to P2P markets were published until 2015, nearly a total of 54 publications was published after 2015. This reveals increased interest from the scientific and industrial communities in P2P electricity markets. Many papers discuss the main drivers, barriers and market designs for P2P electricity markets. Yet, there are other research questions not addressed by the literature, such as life-cycle assessment of hardware and economy of scale, and these other aspects were not extensively covered in this paper.

%% Acknowledgement section
\section*{Acknowledgements}
The work is partly supported by the Danish ForskEL and EUDP programmes through the Energy Collective project (grant no. 2016-1-12530), and by the EU Interreg programme through the Smart City Accelerator project (grant no. 20200999). The post-doctoral grant of Tiago Soares was financed by the ERDF – European Regional Development Fund through the Operational Programme for Competitiveness and Internationalisation - COMPETE 2020 Programme, and by National Funds through the Portuguese funding agency, FCT - Funda\c{c}\~{a}o para a Ci\^{e}ncia e a Tecnologia, within project ESGRIDS - Desenvolvimento Sustentável da Rede Elétrica Inteligente/SAICTPAC/0004/2015-POCI-01-0145-FEDER-016434. The icons used in Figures \ref{Fig_Full_P2P}, \ref{Fig_Full_P2P_maths}, \ref{Fig_Commu_Market}. \ref{Fig_Commu_Market_maths} and \ref{Fig_Hybrid_P2P} were designed by Freepik, Nikita Golubev, Pixel Perfect, Vectors Market, Pixel Buddha, Becris from Flaticon\footnote{\url{https://www.flaticon.com/}}.

%\section*{References}
\bibliographystyle{elsarticle-num}
\bibliography{P2P_ref}

\begin{thebibliography}{10}
\expandafter\ifx\csname url\endcsname\relax
  \def\url#1{\texttt{#1}}\fi
\expandafter\ifx\csname urlprefix\endcsname\relax\def\urlprefix{URL }\fi
\expandafter\ifx\csname href\endcsname\relax
  \def\href#1#2{#2} \def\path#1{#1}\fi

\bibitem{Bussar2016}
C.~Bussar, P.~Stöcker, Z.~Cai, L.~M. Jr., D.~Magnor, P.~Wiernes, N.~van
  Bracht, A.~Moser, D.~U. Sauer, Large-scale integration of renewable energies
  and impact on storage demand in a european renewable power system of
  2050--sensitivity study, Journal of Energy Storage 6 (2016) 1 -- 10.
\newblock \href {http://dx.doi.org/10.1016/j.est.2016.02.004}
  {\path{doi:10.1016/j.est.2016.02.004}}.

\bibitem{Saad2014}
A.~S. al~sumaiti, M.~H. Ahmed, M.~M.~A. Salama, Smart home activities: A
  literature review, Electric Power Components and Systems 42~(3-4) (2014)
  294--305.
\newblock \href {http://dx.doi.org/10.1080/15325008.2013.832439}
  {\path{doi:10.1080/15325008.2013.832439}}.

\bibitem{Zafar2018}
R.~Zafar, A.~Mahmood, S.~Razzaq, W.~Ali, U.~Naeem, K.~Shehzad, Prosumer based
  energy management and sharing in smart grid, Renewable and Sustainable Energy
  Reviews 82 (2018) 1675 -- 1684.
\newblock \href {http://dx.doi.org/https://doi.org/10.1016/j.rser.2017.07.018}
  {\path{doi:https://doi.org/10.1016/j.rser.2017.07.018}}.

\bibitem{Eurelectric2015}
{Eurelectric},
  \href{http://www.elecpor.pt/pdf/18_06_2015_Prosumers_an_integral_part_of_the_power_system_and_market_june.pdf}{Prosumers
  - an integral part of the power system and the market}, June 2015 (Accessed
  on August 2017).
\newline\urlprefix\url{http://www.elecpor.pt/pdf/18_06_2015_Prosumers_an_integral_part_of_the_power_system_and_market_june.pdf}

\bibitem{Schoor2015}
T.~van~der Schoor, B.~Scholtens, Power to the people: Local community
  initiatives and the transition to sustainable energy, Renewable and
  Sustainable Energy Reviews 43 (2015) 666 -- 675.
\newblock \href {http://dx.doi.org/10.1016/j.rser.2014.10.089}
  {\path{doi:10.1016/j.rser.2014.10.089}}.

\bibitem{Bertsch2016}
V.~Bertsch, M.~Hall, C.~Weinhardt, W.~Fichtner, Public acceptance and
  preferences related to renewable energy and grid expansion policy: Empirical
  insights for germany, Energy 114~(Supplement C) (2016) 465 -- 477.
\newblock \href {http://dx.doi.org/10.1016/j.energy.2016.08.022}
  {\path{doi:10.1016/j.energy.2016.08.022}}.

\bibitem{Selloni2017}
D.~Selloni, Codesign for public-interest services, Research for Development,
  Springer, 2017.
\newblock \href {http://dx.doi.org/10.1007/978-3-319-53243-1}
  {\path{doi:10.1007/978-3-319-53243-1}}.

\bibitem{Raworth2017}
K.~Raworth, Doughnut economics: seven ways to think like a 21st-century
  economist, Random house business, 2017.

\bibitem{Bollier2015}
D.~Bollier, Commoning as a transformative social paradigm, The next system
  project, 2015.

\bibitem{Pais2015}
I.~Pais, G.~Provasi, Sharing economy: A step towards the re-embeddedness of the
  economy?, Stato e mercato, Rivista quadrimestrale~(3) (2015) 347--378.
\newblock \href {http://dx.doi.org/10.1425/81604} {\path{doi:10.1425/81604}}.

\bibitem{Hu2017}
J.~Hu, R.~Harmsen, W.~Crijns-Graus, E.~Worrell, M.~van~den Broek, Identifying
  barriers to large-scale integration of variable renewable electricity into
  the electricity market: A literature review of market design, Renewable and
  Sustainable Energy Reviews 81~(Part 2) (2018) 2181 -- 2195.
\newblock \href {http://dx.doi.org/10.1016/j.rser.2017.06.028}
  {\path{doi:10.1016/j.rser.2017.06.028}}.

\bibitem{Peng2017}
D.~Peng, R.~Poudineh, Electricity market design for a decarbonised future: An
  integrated approach, The Oxford Institute for Energy Studies, University of
  Oxford, 2017.
\newblock \href {http://dx.doi.org/10.26889/9781784670948}
  {\path{doi:10.26889/9781784670948}}.

\bibitem{Faber2014}
I.~Faber, W.~Lane, W.~Pak, M.~Prakel, C.~Rocha, J.~V. Farr, Micro-energy
  markets: The role of a consumer preference pricing strategy on microgrid
  energy investment, Energy 74~(Supplement C) (2014) 567 -- 575.
\newblock \href {http://dx.doi.org/10.1016/j.energy.2014.07.022}
  {\path{doi:10.1016/j.energy.2014.07.022}}.

\bibitem{Wu1995}
F.~F. Wu, P.~Varaiya, Coordinated multilateral trades for electric power
  networks: Theory and implementation, Working papers series of the Program on
  Workable Energy Regulation (POWER), June 1995.

\bibitem{Wu1999}
F.~F. Wu, P.~Varaiya, Coordinated multilateral trades for electric power
  networks: Theory and implementation, International Journal of Electrical
  Power \& Energy Systems 21~(2) (1999) 75 -- 102.
\newblock \href {http://dx.doi.org/10.1016/S0142-0615(98)00031-3}
  {\path{doi:10.1016/S0142-0615(98)00031-3}}.

\bibitem{Giotitsas2015}
C.~Giotitsas, A.~Pazaitis, V.~Kostakis, A peer-to-peer approach to energy
  production, Technology in Society 42 (2015) 28 -- 38.
\newblock \href {http://dx.doi.org/10.1016/j.techsoc.2015.02.002}
  {\path{doi:10.1016/j.techsoc.2015.02.002}}.

\bibitem{Morstyn2018b}
T.~Morstyn, N.~Farrell, S.~J. Darby, M.~D. McCulloch, Using peer-to-peer
  energy-trading platforms to incentivize prosumers to form federated power
  plants, Nature Energy 3~(2) (2018) 94--101.
\newblock \href {http://dx.doi.org/10.1038/s41560-017-0075-y}
  {\path{doi:10.1038/s41560-017-0075-y}}.

\bibitem{Schollmeier2001}
R.~Schollmeier, A definition of peer-to-peer networking for the classification
  of peer-to-peer architectures and applications, in: Proceedings First
  International Conference on Peer-to-Peer Computing, 2001, pp. 101--102.
\newblock \href {http://dx.doi.org/10.1109/P2P.2001.990434}
  {\path{doi:10.1109/P2P.2001.990434}}.

\bibitem{Singh2001}
M.~Singh, Peering at peer-to-peer computing, IEEE Internet Computing 5~(6)
  (2001) 4--5.
\newblock \href {http://dx.doi.org/10.1109/MIC.2001.968826}
  {\path{doi:10.1109/MIC.2001.968826}}.

\bibitem{Kant2002}
K.~Kant, R.~Iyer, V.~Tewari, A framework for classifying peer-to-peer
  technologies, in: Cluster Computing and the Grid, 2002. 2nd IEEE/ACM
  International Symposium on, 2002, pp. 368--368.
\newblock \href {http://dx.doi.org/10.1109/CCGRID.2002.1017163}
  {\path{doi:10.1109/CCGRID.2002.1017163}}.

\bibitem{Oram2001}
A.~Oram (Ed.), Peer-to-peer: Harnessing the power of disruptive technologies,
  O'Reilly Media, Sebastopol, CA, USA, 2001.

\bibitem{Aberer2002}
K.~Aberer, M.~Hauswirth, An overview on peer-to-peer information systems, in:
  Proceeding of the workshop on distributed data and structures (WDAS), 2002.

\bibitem{Benkler2006}
Y.~Benkler, The wealth of networks: How social production transforms markets
  and freedom, Yale University Press, 2006.

\bibitem{Vu2010}
Q.~H. Vu, M.~Lupu, B.~C. Ooi, Peer-to-peer computing: Principles and
  applications, 1st Edition, Springer, 2010.
\newblock \href {http://dx.doi.org/10.1007/978-3-642-03514-2}
  {\path{doi:10.1007/978-3-642-03514-2}}.

\bibitem{Peleg2011}
{David Peleg (Ed.)}, Distributed Computing, Vol. 6950 of Theoretical Computer
  Science and General Issues, Springer, 2011.
\newblock \href {http://dx.doi.org/10.1007/978-3-642-24100-0}
  {\path{doi:10.1007/978-3-642-24100-0}}.

\bibitem{Kostakis2016}
V.~Kostakis, A.~Roos, M.~Bauwens, Towards a political ecology of the digital
  economy: Socio-environmental implications of two competing value models,
  Environmental Innovation and Societal Transitions 18 (2016) 82 -- 100.
\newblock \href {http://dx.doi.org/10.1016/j.eist.2015.08.002}
  {\path{doi:10.1016/j.eist.2015.08.002}}.

\bibitem{Einav2016}
L.~Einav, C.~Farronato, J.~Levin, Peer-to-peer markets, Annual Review of
  Economics 8~(1) (2016) 615--635.
\newblock \href {http://dx.doi.org/10.1146/annurev-economics-080315-015334}
  {\path{doi:10.1146/annurev-economics-080315-015334}}.

\bibitem{Beitollahi2007}
H.~Beitollahi, G.~Deconinck, Peer-to-peer networks applied to power grid, in:
  Proceedings of the International Conference on Risks and Security of Internet
  and Systems (CRiSIS), 2007.

\bibitem{Mengelkamp2017}
E.~Mengelkamp, J.~Gärttner, K.~Rock, S.~Kessler, L.~Orsini, C.~Weinhardt,
  Designing microgrid energy markets: A case study: The brooklyn microgrid,
  Applied Energy 105 (2017) 870--880.
\newblock \href {http://dx.doi.org/10.1016/j.apenergy.2017.06.054}
  {\path{doi:10.1016/j.apenergy.2017.06.054}}.

\bibitem{EU2016}
{European commission},
  \href{http://eur-lex.europa.eu/legal-content/EN/TXT/?uri=COM:\\2016:861:FIN}{Proposal
  for a regulation of the European parliament and of the council on the
  internal market for electricity}, 2016, (Accessed on August 2017).
\newline\urlprefix\url{http://eur-lex.europa.eu/legal-content/EN/TXT/?uri=COM:\\2016:861:FIN}

\bibitem{French2017}
{French Government},
  \href{https://www.legifrance.gouv.fr/eli/loi/2017/2/24/DEVR1623346L/jo/texte}{LOI
  n\textsuperscript{o} 2017-227 on self-consumption and renewable energy
  production, JORF no. 48}, February 2017.
\newline\urlprefix\url{https://www.legifrance.gouv.fr/eli/loi/2017/2/24/DEVR1623346L/jo/texte}

\bibitem{Jogunola2017}
O.~Jogunola, A.~Ikpehai, K.~Anoh, B.~Adebisi, M.~Hammoudeh, S.-Y. Son,
  G.~Harris, State-of-the-art and prospects for peer-to-peer transaction-based
  energy system, Energies 10~(12) (2017) 1--28.
\newblock \href {http://dx.doi.org/10.3390/en10122106}
  {\path{doi:10.3390/en10122106}}.

\bibitem{Tushar2018}
W.~Tushar, C.~Yuen, H.~Mohsenian-Rad, T.~Saha, V.~Poor, K.~Wood,
  \href{https://arxiv.org/abs/1804.00962}{Transforming energy networks via peer
  to peer energy trading: Potential of game theoretic approaches}, Under
  review.
\newline\urlprefix\url{https://arxiv.org/abs/1804.00962}

\bibitem{Bower2000}
J.~Bower, D.~W. Bunn, \href{http://www.jstor.org/stable/41322889}{Model-based
  comparisons of pool and bilateral markets for electricity}, The Energy
  Journal 21~(3) (2000) 1--29.
\newline\urlprefix\url{http://www.jstor.org/stable/41322889}

\bibitem{Hausman2008}
E.~Hausman, R.~Hornby, A.~Smith, Bilateral contracting in deregulated
  electricity markets, Report to the American Public Power Association by
  Synapse Energy Economics, Inc, 2008.

\bibitem{Gui2017}
E.~M. Gui, M.~Diesendorf, I.~MacGill, Distributed energy infrastructure
  paradigm: Community microgrids in a new institutional economics context,
  Renewable and Sustainable Energy Reviews 72~(Supplement C) (2017) 1355 --
  1365.
\newblock \href {http://dx.doi.org/10.1016/j.rser.2016.10.047}
  {\path{doi:10.1016/j.rser.2016.10.047}}.

\bibitem{Hirsch2018}
A.~Hirsch, Y.~Parag, J.~Guerrero, Microgrids: A review of technologies, key
  drivers, and outstanding issues, Renewable and Sustainable Energy Reviews 90
  (2018) 402 -- 411.
\newblock \href {http://dx.doi.org/10.1016/j.rser.2018.03.040}
  {\path{doi:10.1016/j.rser.2018.03.040}}.

\bibitem{Eurelectric2013}
{Eurelectric},
  \href{https://www3.eurelectric.org/publications/filtered?pa=1466&page=5}{Active
  distribution system management - a key tool for the smooth integration of
  distributed generation}, February 2013, (Accessed on July 2017).
\newline\urlprefix\url{https://www3.eurelectric.org/publications/filtered?pa=1466&page=5}

\bibitem{Zhao2014}
J.~Zhao, C.~Wang, B.~Zhao, F.~Lin, Q.~Zhou, Y.~Wang, A review of active
  management for distribution networks: Current status and future development
  trends, Electric Power Components and Systems 42~(3-4) (2014) 280--293.
\newblock \href {http://dx.doi.org/10.1080/15325008.2013.862325}
  {\path{doi:10.1080/15325008.2013.862325}}.

\bibitem{Palizban2014}
O.~Palizban, K.~Kauhaniemi, J.~M. Guerrero, Microgrids in active network
  management-part i: Hierarchical control, energy storage, virtual power
  plants, and market participation, Renewable and Sustainable Energy Reviews
  36~(Supplement C) (2014) 428 -- 439.
\newblock \href {http://dx.doi.org/10.1016/j.rser.2014.01.016}
  {\path{doi:10.1016/j.rser.2014.01.016}}.

\bibitem{Liu2017}
N.~Liu, X.~Yu, C.~Wang, C.~Li, L.~Ma, J.~Lei, Energy-sharing model with
  price-based demand response for microgrids of peer-to-peer prosumers, IEEE
  Transactions on Power Systems 32~(5) (2017) 3569--3583.
\newblock \href {http://dx.doi.org/10.1109/TPWRS.2017.2649558}
  {\path{doi:10.1109/TPWRS.2017.2649558}}.

\bibitem{Ilic2012}
D.~Ilic, P.~G.~D. Silva, S.~Karnouskos, M.~Griesemer, An energy market for
  trading electricity in smart grid neighbourhoods, in: 2012 6th IEEE
  International Conference on Digital Ecosystems and Technologies (DEST), 2012,
  pp. 1--6.
\newblock \href {http://dx.doi.org/10.1109/DEST.2012.6227918}
  {\path{doi:10.1109/DEST.2012.6227918}}.

\bibitem{Zhang2017}
C.~Zhang, J.~Wu, C.~Long, M.~Cheng, Review of existing peer-to-peer energy
  trading projects, Energy Procedia 105 (2017) 2563 -- 2568, 8th International
  Conference on Applied Energy, ICAE2016, 8-11 October 2016, Beijing, China.
\newblock \href {http://dx.doi.org/10.1016/j.egypro.2017.03.737}
  {\path{doi:10.1016/j.egypro.2017.03.737}}.

\bibitem{Massague2017}
E.~Bullich-Massagué, M.~Aragüés-Peñalba, P.~Olivella-Rosell,
  P.~Lloret-Gallego, J.~A. Vidal-Clos, A.~Sumper, Architecture definition and
  operation testing of local electricity markets. the empower project, in: 2017
  International Conference on Modern Power Systems (MPS), 2017, pp. 1--5.
\newblock \href {http://dx.doi.org/10.1109/MPS.2017.7974447}
  {\path{doi:10.1109/MPS.2017.7974447}}.

\bibitem{Mihaylov2014}
M.~Mihaylov, S.~Jurado, N.~Avellana, K.~V. Moffaert, I.~M. de~Abril, A.~Nowé,
  {NRGcoin}: Virtual currency for trading of renewable energy in smart grids,
  in: 11th International Conference on the European Energy Market (EEM14),
  2014, pp. 1--6.
\newblock \href {http://dx.doi.org/10.1109/EEM.2014.6861213}
  {\path{doi:10.1109/EEM.2014.6861213}}.

\bibitem{Hasse2017}
F.~Hasse, Paving the way for the energy world of tomorrow, PwC, Berlin - May
  11, 2017, (Accessed on October 2017).

\bibitem{Johnston2017}
J.~Johnston, Chapter 16 - peer-to-peer energy matching: Transparency, choice,
  and locational grid pricing, in: F.~P. Sioshansi (Ed.), Innovation and
  Disruption at the Grid's Edge, Academic Press, 2017, pp. 319 -- 330.
\newblock \href {http://dx.doi.org/10.1016/B978-0-12-811758-3.00016-4}
  {\path{doi:10.1016/B978-0-12-811758-3.00016-4}}.

\bibitem{Parag2016}
Y.~Parag, B.~K. Sovacool, Electricity market design for the prosumer era,
  Nature energy 1 (2016) 16032.
\newblock \href {http://dx.doi.org/10.1038/nenergy.2016.32}
  {\path{doi:10.1038/nenergy.2016.32}}.

\bibitem{Sorin2017}
E.~Sorin, L.~A. Bobo, P.~Pinson, Consensus-based approach to peer-to-peer
  electricity markets with product differentiation, Under review.

\bibitem{Morstyn2018}
T.~Morstyn, A.~Teytelboym, M.~D. McCulloch, Bilateral contract networks for
  peer-to-peer energy trading, IEEE Transactions on Smart Grid PP~(99) (2018)
  1--1.
\newblock \href {http://dx.doi.org/10.1109/TSG.2017.2786668}
  {\path{doi:10.1109/TSG.2017.2786668}}.

\bibitem{Hermana2016}
R.~Alvaro-Hermana, J.~Fraile-Ardanuy, P.~J. Zufiria, L.~Knapen, D.~Janssens,
  Peer to peer energy trading with electric vehicles, IEEE Intelligent
  Transportation Systems Magazine 8~(3) (2016) 33--44.
\newblock \href {http://dx.doi.org/10.1109/MITS.2016.2573178}
  {\path{doi:10.1109/MITS.2016.2573178}}.

\bibitem{Hug2015}
G.~Hug, S.~Kar, C.~Wu, Consensus {+} innovations approach for distributed
  multiagent coordination in a microgrid, IEEE Transactions on Smart Grid 6~(4)
  (2015) 1893--1903.
\newblock \href {http://dx.doi.org/10.1109/TSG.2015.2409053}
  {\path{doi:10.1109/TSG.2015.2409053}}.

\bibitem{Conejo2006}
A.~Conejo, E.~Castillo, R.~Minguez, R.~Garcia-Bertrand, Decomposition
  techniques in mathematical programming: Engineering and science applications,
  Springer, 2006.
\newblock \href {http://dx.doi.org/10.1007/3-540-27686-6}
  {\path{doi:10.1007/3-540-27686-6}}.

\bibitem{Boyd2010}
S.~Boyd, N.~Parikh, E.~Chu, B.~Peleato, J.~Eckstein, Distributed optimization
  and statistical learning via the alternating direction method of multipliers,
  Foundations and Trends in Machine Learning 3~(1) (2011) 1--122.
\newblock \href {http://dx.doi.org/10.1561/2200000016}
  {\path{doi:10.1561/2200000016}}.

\bibitem{Akter2016}
M.~N. Akter, M.~A. Mahmud, A.~M.~T. Oo, A hierarchical transactive energy
  management system for microgrids, in: 2016 IEEE Power and Energy Society
  General Meeting (PESGM), 2016, pp. 1--5.
\newblock \href {http://dx.doi.org/10.1109/PESGM.2016.7741099}
  {\path{doi:10.1109/PESGM.2016.7741099}}.

\bibitem{Rosell2016}
P.~Olivella-Rosell, G.~Viñals-Canal, A.~Sumper, R.~Villafafila-Robles, B.~A.
  Bremdal, I.~Ilieva, S.~. Ottesen, Day-ahead micro-market design for
  distributed energy resources, in: 2016 IEEE International Energy Conference
  (ENERGYCON), 2016, pp. 1--6.
\newblock \href {http://dx.doi.org/10.1109/ENERGYCON.2016.7513961}
  {\path{doi:10.1109/ENERGYCON.2016.7513961}}.

\bibitem{Verschae2016}
R.~Verschae, T.~Kato, T.~Matsuyama, Energy management in prosumer communities:
  A coordinated approach, Energies 9~(7) (2016) 1--27.
\newblock \href {http://dx.doi.org/10.3390/en9070562}
  {\path{doi:10.3390/en9070562}}.

\bibitem{Ilieva2016}
I.~Ilieva, B.~Bremdal, S.~. Ottesen, J.~Rajasekharan, P.~Olivella-Rosell,
  Design characteristics of a smart grid dominated local market, in: CIRED
  Workshop 2016, 2016, pp. 1--4.
\newblock \href {http://dx.doi.org/10.1049/cp.2016.0785}
  {\path{doi:10.1049/cp.2016.0785}}.

\bibitem{Moret2018a}
F.~Moret, P.~Pinson, Energy collectives: A community and fairness based
  approach to future electricity markets, IEEE Transactions on Power Systems
  PP~(99) (2018) 1--1.
\newblock \href {http://dx.doi.org/10.1109/TPWRS.2018.2808961}
  {\path{doi:10.1109/TPWRS.2018.2808961}}.

\bibitem{Morstyn2018c}
T.~Morstyn, M.~McCulloch, Multi-class energy management for peer-to-peer energy
  trading driven by prosumer preferences, IEEE Transactions on Power Systems
  (2018) 1--1\href {http://dx.doi.org/10.1109/TPWRS.2018.2834472}
  {\path{doi:10.1109/TPWRS.2018.2834472}}.

\bibitem{Tushar2016}
W.~Tushar, B.~Chai, C.~Yuen, S.~Huang, D.~B. Smith, H.~V. Poor, Z.~Yang, Energy
  storage sharing in smart grid: A modified auction-based approach, IEEE
  Transactions on Smart Grid 7~(3) (2016) 1462--1475.
\newblock \href {http://dx.doi.org/10.1109/TSG.2015.2512267}
  {\path{doi:10.1109/TSG.2015.2512267}}.

\bibitem{Long2017}
C.~Long, J.~Wu, C.~Zhang, M.~Cheng, A.~Al-Wakeel, Feasibility of peer-to-peer
  energy trading in low voltage electrical distribution networks, Energy
  Procedia 105 (2017) 2227 -- 2232, 8th International Conference on Applied
  Energy, ICAE2016, 8-11 October 2016, Beijing, China.
\newblock \href {http://dx.doi.org/10.1016/j.egypro.2017.03.632}
  {\path{doi:10.1016/j.egypro.2017.03.632}}.

\bibitem{Liu2015}
T.~Liu, X.~Tan, B.~Sun, Y.~Wu, X.~Guan, D.~H.~K. Tsang, Energy management of
  cooperative microgrids with p2p energy sharing in distribution networks, in:
  2015 IEEE International Conference on Smart Grid Communications
  (SmartGridComm), 2015, pp. 410--415.
\newblock \href {http://dx.doi.org/10.1109/SmartGridComm.2015.7436335}
  {\path{doi:10.1109/SmartGridComm.2015.7436335}}.

\bibitem{Sorin2017b}
E.~Sorin, Peer-to-peer electricity markets with product differentiation - large
  scale impact of a consumer-oriented market, Master thesis in Technical
  University of Denmark,, 2017.

\bibitem{Kang2017}
J.~Kang, R.~Yu, X.~Huang, S.~Maharjan, Y.~Zhang, E.~Hossain, Enabling localized
  peer-to-peer electricity trading among plug-in hybrid electric vehicles using
  consortium blockchains, IEEE Transactions on Industrial Informatics PP~(99)
  (2017) 1--1.
\newblock \href {http://dx.doi.org/10.1109/TII.2017.2709784}
  {\path{doi:10.1109/TII.2017.2709784}}.

\bibitem{Moret2018b}
F.~Moret, T.~Baroche, E.~Sorin, P.~Pinson,
  \href{http://pierrepinson.com/docs/Moretetal17PSCC.pdf}{Negotiation
  algorithms for peer-to-peer electricity markets: Computational properties},
  in: Accepted on 20th Power System Computation Confereence, PSCC, 2018.
\newline\urlprefix\url{http://pierrepinson.com/docs/Moretetal17PSCC.pdf}

\bibitem{Sikorski2017}
J.~J. Sikorski, J.~Haughton, M.~Kraft, Blockchain technology in the chemical
  industry: Machine-to-machine electricity market, Applied Energy 195 (2017)
  234--246.
\newblock \href {http://dx.doi.org/10.1016/j.apenergy.2017.03.039}
  {\path{doi:10.1016/j.apenergy.2017.03.039}}.

\bibitem{Drescher2017}
D.~Drescher, Blockchain basics: A non-technical introduction in 25 steps,
  Apress, 2017.
\newblock \href {http://dx.doi.org/10.1007/978-1-4842-2604-9}
  {\path{doi:10.1007/978-1-4842-2604-9}}.

\bibitem{PWC2016}
{PricewaterhouseCoopers},
  \href{https://www.pwc.com/gx/en/industries/assets/pwc-blockchain-opportunity-for-energy-producers-and-consumers.pdf}{Blockchain
  - an opportunity for energy producers and consumers?}, 2016, (Accessed on
  October 2017).
\newline\urlprefix\url{https://www.pwc.com/gx/en/industries/assets/pwc-blockchain-opportunity-for-energy-producers-and-consumers.pdf}

\bibitem{Andoni2017}
M.~Andoni, V.~Robu, D.~Flynn, Crypto-control your own energy supply, Nature
  548~(7666) (2017) 158.
\newblock \href {http://dx.doi.org/10.1038/548158b}
  {\path{doi:10.1038/548158b}}.

\bibitem{Vangulick2018}
D.~Vangulick, B.~Cornélusse, E.~Damien,
  \href{http://hdl.handle.net/2268/220759}{Blockchain for peer-to-peer energy
  exchanges: Design and recommendations}, Under review.
\newline\urlprefix\url{http://hdl.handle.net/2268/220759}

\bibitem{Bozic2016}
N.~Bozic, G.~Pujolle, S.~Secci, A tutorial on blockchain and applications to
  secure network control-planes, in: 3rd Smart Cloud Networks Systems (SCNS),
  2016, pp. 1--8.
\newblock \href {http://dx.doi.org/10.1109/SCNS.2016.7870552}
  {\path{doi:10.1109/SCNS.2016.7870552}}.

\bibitem{Brudermann2014}
T.~Brudermann, Y.~Yamagata, Towards an agent-based model of urban electricity
  sharing, in: International Conference and Utility Exhibition on Green Energy
  for Sustainable Development (ICUE), 2014, pp. 1--5.

\bibitem{Ploug2017}
R.~Jenle, T.~Pallesen, How engineers make markets organizing electricity system
  decarbonization, Revue Francaise de Sociologie 58~(3) (2017) 375--397.

\bibitem{Pallesen2018}
T.~Pallesen, R.~P. Jenle, Organizing consumers for a decarbonized electricity
  system: Calculative agencies and user scripts in a danish demonstration
  project, Energy Research \& Social Science 38 (2018) 102--109.
\newblock \href {http://dx.doi.org/10.1016/j.erss.2018.02.003}
  {\path{doi:10.1016/j.erss.2018.02.003}}.

\bibitem{Gyamfi2013}
S.~Gyamfi, S.~Krumdieck, T.~Urmee, Residential peak electricity demand
  response—highlights of some behavioural issues, Renewable and Sustainable
  Energy Reviews 25 (2013) 71 -- 77.
\newblock \href {http://dx.doi.org/10.1016/j.rser.2013.04.006}
  {\path{doi:10.1016/j.rser.2013.04.006}}.

\bibitem{Allcott2010}
H.~Allcott, S.~Mullainathan, Behavior and energy policy, Science 327~(5970)
  (2010) 1204--1205.
\newblock \href {http://dx.doi.org/10.1126/science.1180775}
  {\path{doi:10.1126/science.1180775}}.

\bibitem{Simon1997}
H.~A. Simon, An empirically based microeconomics, Cambridge University Press
  Cambridge, U.K, 1997.

\bibitem{Baroche2018}
T.~Baroche, P.~Pinson, R.~Le~Goff~Latimier, H.~Ben~Ahmed,
  \href{https://arxiv.org/abs/1803.02159}{Exogenous approach to grid cost
  allocation in peer-to-peer electricity markets}, Under review.
\newline\urlprefix\url{https://arxiv.org/abs/1803.02159}

\bibitem{Guerrero2018}
J.~Guerrero, A.~Chapman, G.~Verbic,
  \href{https://arxiv.org/abs/1809.06976}{Decentralized p2p energy trading
  under network constraints in a low-voltage network}, Under review.
\newline\urlprefix\url{https://arxiv.org/abs/1809.06976}

\bibitem{Kargarian2017}
A.~Kargarian, J.~Mohammadi, J.~Guo, S.~Chakrabarti, M.~Barati, G.~Hug, S.~Kar,
  R.~Baldick, Toward distributed/decentralized dc optimal power flow
  implementation in future electric power systems, IEEE Transactions on Smart
  Grid PP~(99) (2017) 1--1.
\newblock \href {http://dx.doi.org/10.1109/TSG.2016.2614904}
  {\path{doi:10.1109/TSG.2016.2614904}}.

\bibitem{IEEE_14_bus}
\href{https://www2.ee.washington.edu/research/pstca/pf14/pg_tca14bus.htm}{Archive
  of ieee 14-bus network system}.
\newline\urlprefix\url{https://www2.ee.washington.edu/research/pstca/pf14/pg_tca14bus.htm}

\bibitem{Dowell2016}
J.~Dowell, P.~Pinson, Very-short-term probabilistic wind power forecasts by
  sparse vector autoregression, IEEE Transactions on Smart Grid 7~(2) (2016)
  763--770.
\newblock \href {http://dx.doi.org/10.1109/TSG.2015.2424078}
  {\path{doi:10.1109/TSG.2015.2424078}}.

\bibitem{Ratnam2017}
E.~L. Ratnam, S.~R. Weller, C.~M. Kellett, A.~T. Murray, Residential load and
  rooftop pv generation: An australian distribution network dataset,
  International Journal of Sustainable Energy 36~(8) (2017) 787--806.
\newblock \href {http://dx.doi.org/10.1080/14786451.2015.1100196}
  {\path{doi:10.1080/14786451.2015.1100196}}.

\bibitem{P2P_test_case2018}
T.~Sousa, T.~Soares, P.~Pinson, F.~Moret, T.~Baroche, E.~Sorin, The p2p-ieee 14
  bus system [data set] (Apr. 2018).
\newblock \href {http://dx.doi.org/10.5281/zenodo.1220935}
  {\path{doi:10.5281/zenodo.1220935}}.

\bibitem{Webster2002}
J.~Webster, R.~T. Watson, Analyzing the past to prepare for the future: Writing
  a literature review, MIS Q. 26~(2) (2002) xiii--xxiii.

\bibitem{Brocke2009}
J.~vom Brocke, A.~Simons, B.~Niehaves, B.~Niehaves, K.~Riemer, R.~Plattfaut,
  A.~Cleven, Reconstructing the giant: On the importance of rigour in
  documenting the literature search process, in: Information systems in a
  globalising world : challenges, ethics and practices ; ECIS 2009, 17th
  European Conference on Information Systems, Universit{\`a} di Verona,
  Facolt{\`a} di Economia, Departimento de Economia Aziendale, Verona, 2009,
  pp. 2206--2217.

\end{thebibliography}


%\begin{thebibliography}{00}

%\bibliographystyle{elsarticle-num.bst}
%\bibliography{myRef}

%\end{thebibliography}
\end{document}